\documentclass[aps,floatfix,amsmath,amssymb,nofootinbib,twocolumn]{revtex4-1}
\pdfoutput=1
\usepackage{graphicx}
\usepackage{dcolumn}
\usepackage{bbm}
\usepackage{epstopdf}
\usepackage[bookmarks=true,colorlinks,linkcolor=blue,urlcolor=blue,citecolor=blue]{hyperref}
\usepackage{bbold}
\usepackage{bm}
\usepackage{color}
\usepackage{subcaption} 
\SetSymbolFont{largesymbols}{bold}{OMX}{txex}{b}{n}

\usepackage{setspace}
\newcommand{\beq}{\begin{equation}}
\newcommand{\eeq}{\end{equation}}
\newcommand{\beqn}{\begin{eqnarray}}
\newcommand{\eeqn}{\end{eqnarray}}
\def\bal#1\eal{\begin{align}#1\end{align}}
\newcommand{\nn}{\nonumber\\}
\newcommand{\eulergamma}[1]{\Gamma \left( #1 \right)}
\newcommand{\bmat}{\begin{pmatrix}}
\newcommand{\emat}{\end{pmatrix}}
\newcommand{\sgn}[1]{\textnormal{sgn}\left( #1 \right)}
\newcommand{\av}[1]{\left\vert #1 \right\vert}

\def \q{{\mathbf{q}}}

\def \k{{\mathbf{k}}}

\def \q{{\mathbf{q}}}

\def \p{{\mathbf{p}}}

\def \psib{{\overline{\psi}}}
\def \Psib{{\overline{\Psi}}}
\def \tk{{\tilde{\mathbf{k}}}}
\def \tq{{\tilde{\mathbf{q}}}}
\def \tp{{\tilde{\mathbf{p}}}}

\def \tn{\textnormal}

\begin{document}

\title{Quenched disorder at antiferromagnetic quantum critical points in $2d$ metals}

\author{Johannes Halbinger}
\author{Matthias Punk}
\affiliation{Physics Department, Arnold Sommerfeld Center for Theoretical Physics, Center for NanoScience, and Munich Center for Quantum Science and Technology (MCQST), Ludwig-Maximilians University Munich, Germany}

\date{\today}

\begin{abstract}

We study spin density wave quantum critical points in two dimensional metals with a quenched disorder potential coupling to the electron density. 
Adopting an $\epsilon$-expansion around three spatial dimensions, where both disorder and the Yukawa-type interaction between electrons and bosonic order parameter fluctuations are marginal, we present a perturbative, one-loop renormalization group analysis of this problem, where the interplay between fermionic and bosonic excitations is fully incorporated.
Considering two different Gaussian disorder models restricted to small-angle scattering, we show that the non-Fermi liquid fixed point of the clean SDW hot-spot model is generically unstable and the theory flows to strong coupling due to a mutual enhancement of interactions and disorder. 
We study properties of the asymptotic flow towards strong coupling, where our perturbative approach eventually breaks down. Our results indicate that disorder dominates at low energies, suggesting that the ground-state in two dimensions is Anderson-localized.

\end{abstract}

\maketitle

\section{Introduction}

In the vicinity of a quantum critical point (QCP) in quasi two-dimensional metals the strong interaction between electrons and order parameter fluctuations destroys the quasiparticle character of electronic excitations, which lies at the heart of Landau's Fermi liquid theory. Indeed, non-Fermi liquids or strange metals are frequently found in the vicinity of magnetic QCPs \cite{Lohneysen2007}, with heavy fermion materials, iron-pnictides and cuprates as prominent examples \cite{Stewart2001,Shibauchi2014,Greene2020}. One striking aspect of such non-Fermi liquid behavior is the observed linear temperature dependence of the electrical resistivity, together with the absence of resistivity saturation at the Mott-Ioffe-Regel bound at high temperatures. The latter is often attributed to a breakdown of the quasi-particle picture in such materials \cite{Emery1995}. 

While the theoretical description of metallic quantum critical points has seen substantial progress using both numerical as well as field-theoretical developments \cite{Berg2018, Lee2018}, central questions still remain open. In particular, the computation of dc transport properties and the question if a linear-in-T resistivity is a generic feature of metallic QCPs poses a formidable challenge, as it requires to account for processes which relax the total momentum of a current carrying state, either via disorder or umklapp scattering. Previous analytic works studied dc transport either using the Boltzmann equation \cite{HlubinaRice1995, Rosch1999, Maslov2011}, or perturbatively in the presence of weak disorder, assuming that disorder doesn't drive the system away from the clean non-Fermi liquid fixed point \cite{Hartnoll 2014, Patel2014, Freire2017}. In this article we investigate if the latter assumption is indeed valid and perform a detailed renormalization group (RG) study of spin-density wave (SDW) quantum critical points in the presence of quenched disorder.

So far, the fate of metallic QCPs with quenched disorder has been primarily studied using the Hertz approach \cite{Hertz1976}, with few notable exceptions \cite{Nosov2020}. In the former the fermionic degrees of freedom are integrated out and one is left with a Ginzburg-Landau-type theory for the overdamped bosonic order parameter fluctuations. Within this approach several possible scenarios about the fate of various quantum critical points in the presence of disorder have been put forward, including the potential for novel fixed points at finite or infinite disorder \cite{Kirkpatrick1996, Narayanan1999, Millis2002, Hoyos2007}. Interestingly, for the SDW QCP in metals with Ising symmetry, it has been argued that the critical point is washed out and replaced by a smooth crossover due to Griffiths effects, i.e.~rare ordered regions in the disorder potential \cite{Vojta2003}. Even though many questions still remain open, it is important to emphasize that the Hertz approach is invalid for clean two-dimensional systems, because the interaction between electrons and bosonic order parameter fluctuations leads to a loss of electronic quasiparticle coherence and to a singular nature of the bosonic $n$-point vertices in the Ginzburg-Landau expansion. Accounting for this delicate interplay requires a careful theoretical analysis treating both fermionic and bosonic excitations on equal footing, which is particularly challenging in a renormalization group setting. Early studies of these models without disorder were based on diagrammatic approaches \cite{Altshuler1994, Kim1994, Polchinski1994, Metzner2003} and were believed to be justified within a large-$N_f$ expansion, with $N_f$ the number of fermionic flavours \cite{Abanov2000}. More recent works showed that such expansions are in fact uncontrolled and these theories remain strongly coupled in the $N_f \to \infty$ limit \cite{Lee2009, Metlitski2010a, Metlitski2010b}.

The main goal of this work is to perform a controlled renormalization group study of antiferromagnetic QCPs in two-dimensional metals in the presence of quenched disorder, by reformulating the epsilon-expansion for the SDW hot-spot model developed by Sur and Lee \cite{Sur2015} on the closed-time Schwinger-Keldysh contour. This allows us to directly perform a disorder average of the partition function and to study the interplay between fermionic excitations, bosonic order parameter fluctuations and disorder induced interactions between the fermions in a controlled manner. In contrast to the above mentioned Hertz-type theories as well as Finkel'stein-type RG studies of interacting electrons with quenched disorder \cite{Finkelshtein, Castellani, Chamon1999, Nosov2020}, the crucial difference in our approach is that electronic degrees of freedom are \emph{not} integrated out and are treated on equal footing with the bosonic excitations. 

Importantly, we only consider models of disorder potentials which impart a small momentum transfer $|\mathbf{p}| \ll k_F$ onto the electrons, with $k_F$ the characteristic Fermi momentum, such that electrons stay in the vicinity of a hot-spot when scattering off the disorder potential. Besides simplifying the analysis, this restriction also ensures that the hot-spot model remains valid, as cold electrons far away from the hot-spots cannot become hot by scattering off the disorder potential with a large momentum transfer and end up in the vicinity of a hot-spot. 

Even though disorder does not lead to backscattering of electrons in our model, we nonetheless expect a non-divergent dc conductivity due to the interplay between disorder and the Yukawa-type interaction of electrons with order parameter fluctuations. Since the latter gives rise to large angle scattering, their combination allows electrons to equilibrate across all hot-spots around the Fermi surface. Moreover, disorder localizes the cold electrons in $2d$ \cite{AbrahamsRamakrishnan1979}, which consequently do not participate in transport and cannot short circuit the contribution of hot electrons as would be the case without disorder \cite{HlubinaRice1995}. 
	
	\begin{figure}
	\centering
	\includegraphics[width=0.9 \columnwidth]{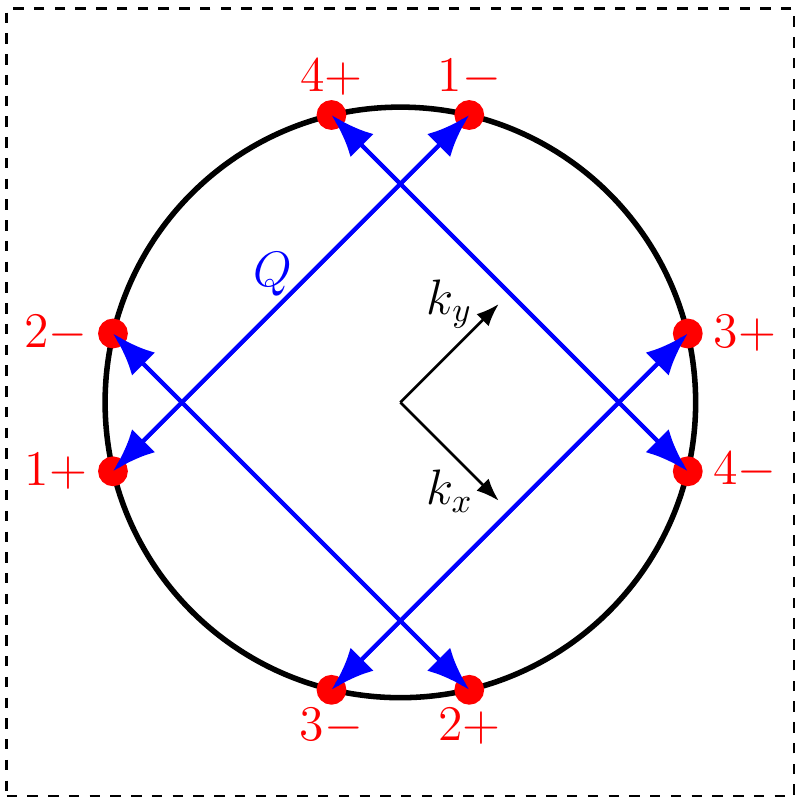}
	\caption{Fermi surface and scattering geometry of electrons near the eight hot-spots connected by the antiferromagnetic ordering wave vector $\mathbf{Q}=(\pi,\pi)$ modulo reciprocal lattice vectors. The dashed line marks the boundary of the 1st Brillouin zone for the square lattice. The momenta $k_x$ and $k_y$ indicate the local coordinate system at the hot-spots which is used in the following.}
	\label{figfermisurface}
	\end{figure}

In this work we focus on two Gaussian disorder models, a $\delta$-correlated disorder potential within single hot-spots, as well as a power-law correlated disorder potential. A simple tree-level scaling analysis seems to indicate that $\delta$-correlated disorder is irrelevant in $d>2$ spatial dimensions, but we show that the linearised electron dispersion in the hot-spot model leads to UV/IR mixing and the effective disorder coupling is in fact marginal in $d=3$, as is the Yukawa interaction. For this reason we can treat both interactions simultaneously in a perturbative epsilon-expansion around three spatial dimensions. Lastly, we refrain from a discussion of potentially important Griffiths effects due to a random boson mass term, as this is beyond the scope of this work.

The remainder of this article is structured as follows: in Sec.~\ref{sec2} we reformulate the hot-spot model of Sur and Lee on the Keldysh contour and discuss our disorder model. Our one-loop RG results for both disorder models are discussed in Sec.~\ref{sec3}, with technical details shifted to the appendices. We conclude with a brief discussion in Sec.~\ref{sec4}.

\section{Model}
\label{sec2}

\subsection{Keldysh formulation of the Sur-Lee model}

We start from a hot-spot model of electrons on the square lattice in two dimensions, coupled to fluctuations of an antiferromagnetic order parameter with ordering wave vector $\mathbf{Q}=(\pi,\pi)$ (see Fig.~\ref{figfermisurface} for a sketch of the hot-spot geometry). In particular we use an extension of this model developed by Sur and Lee, which is amenable to dimensional regularisation \cite{Sur2015}. In their approach the Fermi surface geometry of the $2d$ model is left unchanged, but the number of dimensions orthogonal to the Fermi surface (i.e.~its co-dimension) is increased. A tree-level scaling analysis then shows that a controlled epsilon-expansion around $d=3$ dimensions is possible. Since we are going to study this model in the presence of quenched disorder, we start by reformulating the action of the Sur-Lee model on the closed-time Schwinger-Keldysh contour, as this allows us to directly perform a disorder average of the partition function later on. After Keldysh rotation of the fields its action takes the form (see also Ref.~\cite{PatelStrackSachdev} for a Keldysh formulation of the hot-spot model in two dimensions) 
	\bal
		S= S_\Psi + S_\phi + S_Y + S_{\phi^4}
	\eal
with
	\begin{widetext}
	\bal
		S_\Psi&= \sum_{n,j,\sigma} \int_k \bmat \Psib^{(1)}_{n,j,\sigma}(k) & \Psib^{(2)}_{n,j,\sigma}(k) \emat \bmat \sigma_y (\omega+i\delta) - i \boldsymbol{\sigma}_{\mathbf{z}} \tk - i \sigma_x \varepsilon_n(\k) & \sigma_y \delta_f(\omega) \\ 0 & \sigma_y (\omega-i\delta) - i \boldsymbol{\sigma}_{\mathbf{z}} \tk - i \sigma_x \varepsilon_n(\k) \emat \bmat \Psi^{(1)}_{n,j,\sigma}(k) \\ \Psi^{(2)}_{n,j,\sigma}(k) \emat, \\
		S_\phi&=\frac{1}{2} \int_p \bmat \vec{\phi}^c(p) && \vec{\phi}^q(p) \emat \bmat 0 && 2\left[ \left( \omega-i\delta \right)^2-a^2\tp^2-c^2 \p^2 \right] \\ 2\left[ \left( \omega+i\delta \right)^2-a^2\tp^2-c^2\p^2 \right] && \delta_b(\omega) \emat \bmat \vec{\phi}^c(-p) \\ \vec{\phi}^q(-p) \emat, \\
	S_Y&=- i\frac{g}{\sqrt{N_f}} \sum_{n,j,\sigma,\sigma'}
	\int_{k,p} \Psib^{(a)}_{n,j,\sigma}(k+p)\ \Phi^{\alpha}_{\sigma \sigma'}(p) \gamma^\alpha_{(ab)} \sigma_x\ \Psi^{(b)}_{\bar{n},j,\sigma'}(k), \\
	\begin{split}
	S_{\phi^4}&=-\frac{u_1}{4} \int_{p_1,p_2,p_3} \tn{Tr} \left\{ \Phi^c (p_1+p_3) \Phi^c(p_2-p_3) + \Phi^q (p_1+p_3) \Phi^q(p_2-p_3) \right\} \tn{Tr} \left\{ \Phi^c (-p_1) \Phi^q(-p_2) \right\} \\
	&-\frac{u_2}{4} \int_{p_1,p_2,p_3}  \tn{Tr} \left\{ \left[ \Phi^c (p_1+p_3) \Phi^c(p_2-p_3) + \Phi^q (p_1+p_3) \Phi^q(p_2-p_3)\right] \Phi^c (-p_1) \Phi^q(-p_2) \right\}.
	\end{split}
	\label{generaldaction}
	\eal
	\end{widetext}
Here we follow the notation of Sur and Lee where $\Psi^{(i)}_{1,j,\sigma}=\big( \psi^{(i)}_{1+,j,\sigma},\psi^{(i)}_{3+,j,\sigma} \big)^T$, $\Psi^{(i)}_{2,j,\sigma}=\big( \psi^{(i)}_{2+,j,\sigma},\psi^{(i)}_{4+,j,\sigma} \big)^T$, $\Psi^{(i)}_{3,j,\sigma}=\big( \psi^{(i)}_{1-,j,\sigma}, -\psi^{(i)}_{3-,j,\sigma} \big)^T$ and $\Psi^{(i)}_{4,j,\sigma}=\big( \psi^{(i)}_{2-,j,\sigma}, -\psi^{(i)}_{4-,j,\sigma} \big)^T$ are two component fermionic spinors comprised of fermion fields at anti-podal hot-spots (see Fig.~\ref{figfermisurface}), $\Psib = \Psi^\dagger \sigma_y$ with $\sigma_{x,y,z}$ the Pauli matrices, $\boldsymbol{\sigma}_{\mathbf{z}}$ is a $(d-2)$-dimensional vector with entries $\sigma_z$ and the index $n$ runs over the four distinct hot-spot pairs with $\bar{1} = 3$, $\bar{2} = 4$, $\bar{3} = 1$ and $\bar{4} = 2$. The superscript $(i)$ distinguishes the two Keldysh components and the infinitesimal $\delta_f(\omega),\delta_b(\omega) \sim \delta = 0^+$ are needed for convergence. As in Ref.~\cite{Sur2015} the action is generalised to a $SU(N_f)$ fermion flavour group and to a $SU(N_c)$ spin group, with $j$ and $\sigma$ as respective indices (the physical model corresponds to $N_f=1$ and $N_c=2$). SDW order parameter fluctuations are described by the matrix valued field $\Phi^\alpha =  \vec{\phi}^\alpha \cdot \vec{\tau}$, with $\vec{\tau}$ denoting the $N_c^2-1$ generators of $SU(N_c)$ and 
$\alpha \in \{ q, c \} $ is the Keldysh index. The Keldysh rotation was performed using the convention
	\begin{equation}
	\begin{alignedat}{2}
		\vec{\phi}^+ &= \vec{\phi}^c+\vec{\phi}^q, \qquad &&\vec{\phi}^- = \vec{\phi}^c-\vec{\phi}^q,\\
		\psi^+ &= \frac{1}{\sqrt{2}} (\psi^{(1)} + \psi^{(2)}), \qquad &&\psib^+ = \frac{1}{\sqrt{2}} (\psib^{(1)} + \psib^{(2)}), \\
		\psi^- &= \frac{1}{\sqrt{2}} (\psi^{(1)} - \psi^{(2)}),\qquad &&\psib^- = \frac{1}{\sqrt{2}} (\psib^{(2)} - \psib^{(1)})
	\end{alignedat}
	\end{equation}
where the indices $\pm$ denote fields on the forward and backward branch of the closed time contour. Furthermore we defined the linearised fermionic dispersions near the hot-spots 
	\begin{equation}
	\begin{alignedat}{2}
		\varepsilon_1(\k) &= v k_x+k_y, \quad &&\varepsilon_3(\k)= v k_x-k_y, \\
		\varepsilon_2(\k) &= - k_x+v k_y,\quad &&\varepsilon_4(\k)= k_x+ vk_y.
		\label{fermiondisp}
	\end{alignedat}
	\end{equation}
The $(d+1)$-dimensional integral $\int_k = \int\frac{d\omega d\tk d\k}{(2\pi)^{d+1}}$ runs over real frequency $\omega$, the two dimensional momentum $\mathbf{k}=(k_x,k_y)$ as well as the $(d-2)$ additional momentum directions $\tilde{\mathbf{k}}$ orthogonal to the Fermi surface in the $(k_x,k_y)$-plane.

The Yukawa coupling term $\sim g$ between fermions and order parameter fluctuations has been written in compact form with the help of the matrices
	\bal
		\gamma^c_{(ab)} = \bmat 1&0\\ 0&1 \emat,\quad \gamma^q_{(ab)} = \bmat 0&1\\ 1&0 \emat.
	\eal
Note that we use the convention of Greek superscripts for the bosonic Keldysh indices and Latin superscripts in brackets for fermionic Keldysh indices, where a sum over repeated indices is implied.

The bare fermion propagators in general dimensions are given by
	\bal
		-i \left\langle \Psi^{(a)}_{n,j,\sigma}(k) \Psib^{(b)}_{n,j,\sigma}(k) \right\rangle = \bmat G^R_n(k) &G^K_n(k) \\ 0&G^A_n(k) \emat,
	\eal
where the retarded/advanced and Keldysh Green's functions
	\bal
		G^{R/A}_n(k) &= \frac{\sigma_y(\omega \pm i\delta) - i \boldsymbol{\sigma}_{\mathbf{z}} \tk - i \sigma_x \varepsilon_n(\k)}{(\omega \pm i\delta)^2-\tk^2-\varepsilon_n^2(\k)}, \\
		G^{K}_n(k) &= F_f(\omega) \left[ G^R_n(k) - G^A_n(k) \right]
	\eal
are now matrices in spinor space and $F_f(\omega)=\sgn{\omega}$ in thermal equilibrium and at zero temperature. The retarded and advanced propagators are related via $\left( \sigma_y G^R_n \right)^\dagger = \sigma_y G^A_n$ and therefore $\left( \sigma_y G^K_n \right)^\dagger = -\sigma_y G^K_n$. Note that these relations hold for the inverse propagators as well.

For the bare boson propagators we find
	\bal
		-i \left\langle \phi^{\alpha,a}(p) \phi^{\beta,b}(-p) \right\rangle = \bmat D^K(p) &D^R(p) \\ D^A(p)&0 \emat \delta_{ab}
	\eal
($a$ and $b$ labelling the vector component) with
	\bal
		D^{R/A}(p) &=\frac{1}{2} \frac{1}{(\omega \pm i\delta)^2-a^2\tp^2-c^2\p^2},\\
		D^K(p) &= F_b(\omega) \left[ D^R(p)-D^A(p) \right]
	\eal
and $F_b(\omega)=\sgn{\omega}$ in thermal equilibrium and at zero temperature.  The retarded and advanced propagators can be obtained from each other via $\left[ D^R(\omega,\vec{p}) \right]^\dagger = D^A(\omega,\vec{p}) = D^R(-\omega,\vec{p})$, where we used the shorthand notation $\vec{p} = (\tp,\p)$. Consequently, the Keldysh propagator obeys $\left[ D^K(\omega,\vec{p}) \right]^\dagger = - D^K(\omega,\vec{p}) = - D^K(-\omega,\vec{p})$. Again, these relations hold for the inverse propagators as well.

A simple tree level scaling analysis of the action in Eq.~\eqref{generaldaction} in $d+1$ dimensions shows that scaling dimensions of the fields and couplings at the Gaussian fixed point are given by 
	\begin{equation}
	\begin{alignedat}{2}
		&\big[ \Psi \big] = -\frac{d+2}{2}, \qquad &&\big[ \vec{\phi} \big] = -\frac{d+3}{2}, \\
		&\big[ g \big] = \frac{3-d}{2}, &&\big[ u_{1,2} \big] = 3-d,
	\end{alignedat}
	\end{equation}
which shows that both the Yukawa interaction $g$ as well as the $\phi^4$ interactions $u_{1,2}$ are irrelevant in $d>3$ spatial dimensions and an expansion in $\varepsilon = 3-d$ is feasible \cite{Sur2015}.

\subsection{Quenched disorder}

The main goal of this work is to study the model in Eq.~\eqref{generaldaction} in the presence of a quenched, random disorder potential $V$, which couples to the density of fermions via the following term in the action
	\bal
	S_{\text{dis},0} = - \int_{\omega,\vec{k}, \vec{p}} V(\vec{p})\ \Psib^{(a)}_{n,j,\sigma}(\vec{k}+\vec{p},\omega) \sigma_y \Psi^{(a)}_{n,j,\sigma}(\vec{k},\omega)
	\label{Sdis0}
	\eal
with $\vec{k} = (\tk,\k)$ and a sum over repeated indices is implied. To keep the hot-spot model valid and the complexity manageable, we assumed here that the disorder potential scatters fermions only within a single hot-spot, as can be seen from the presence of the $\sigma_y$ matrix and the fact that the two spinors carry the same hot-spot index $n$ in Eq.~\eqref{Sdis0}. Taking an Edwards model with identical, randomly placed impurities as an example for the disorder potential, this would correspond to an impurity scattering potential with a finite range, such that the momentum transfer $\vec{p}$ is much smaller than the characteristic Fermi momentum $|\vec{p}| \ll k_F$.  
This simplifying assumption also precludes cold electrons from becoming hot by scattering off impurities from cold regions of the Fermi surface to hot-spot regions. Only in this case the hot-spot model remains valid and cold electrons away from the hot-spots can be safely omitted from our analysis.

Even though the assumption of a small momentum transfer precludes backscattering due to disorder, it is important to realize that the Yukawa coupling gives rise to such large-angle scattering processes. As already mentioned in the introduction, the combination of disorder and magnetic scattering is thus expected to lead to a non-diverging dc conductivity in this model. 
The small-angle scattering assumption implicit in Eq.~\eqref{Sdis0} drastically simplifies the following analysis, because otherwise the disorder averaged interaction would contain a vast number of different couplings from various hot-spot combinations which renormalise differently. In our model there is only one disorder coupling constant, however.

In this work we study two different Gaussian disorder models with zero mean $\langle V(\vec{p}) \rangle_\text{dis} = 0$ and disorder correlation function
\bal
\langle V(\vec{p}) V(-\vec{p}) \rangle_\text{dis} &= 2 u(\vec{p}).
\eal 
 The first model is a white noise model with $\delta$-correlated disorder in real-space, which is restricted to single hot-spots, whereas the second model exhibits power-law correlated disorder in the two physical dimensions and $\delta$-correlations in the additional $d-2$ dimensions:
\bal
 \delta\text{-correlated:} \hspace{1cm} u(\vec{p}) &= u_0 ,  \label{Eqdisordermodels1} \\
  \text{power-law:} \hspace{1cm} u(\vec{p}) &= u_0 \, |\mathbf{p}|^{\alpha-2}
  \label{Eqdisordermodels2}
 \eal
with exponent $\alpha>0$. Note again that our $\delta$-correlated disorder model is different from a standard white noise model with arbitrary large momentum transfer, as electrons only scatter within the same hot-spot in our case.

The Keldysh formalism allows us to directly perform a disorder average of the partition function over all possible realisations of the Gaussian distributed potential $V(\vec{p})$, with this disorder average defined by
	\bal
		\langle \cdots \rangle_\text{dis} = \int \mathcal{D}[V(\vec{p})] \, \cdots \, \exp\left\{-\frac{1}{4} \int_{\vec{p}} \frac{V(\vec{p}) \, V(-\vec{p})}{u(\vec{p})} \right\}.
	\eal
This leads to the following additional, disorder induced four-fermion interaction term in the action
	\begin{widetext}
	\bal
	S_\text{dis} =	i\sum_{n,n'}\sum_{j,j'} \sum_{\sigma,\sigma'} \int_{\overset{\vec{k}_1,\vec{k}_2,\vec{k}_3}{\omega_1,\omega_2}} \left[ \Psib^{(a)}_{n,j,\sigma}(\vec{k}_1+\vec{k}_3,\omega_1) \sigma_y \Psi^{(a)}_{n,j,\sigma}(\vec{k}_1,\omega_1) \right] u(\vec{k}_3) \left[ \Psib^{(b)}_{n',j',\sigma'}(\vec{k}_2-\vec{k}_3,\omega_2) \sigma_y \Psi^{(b)}_{n',j',\sigma'}(\vec{k}_2,\omega_2) \right]
	\eal
	\end{widetext}
with $u(\vec{k}_3)$ given in Eqs.~\eqref{Eqdisordermodels1} and \eqref{Eqdisordermodels2}. Note that $u_0$ is restricted to positive values for the functional integral to be well defined.
The tree level scaling dimension of the disorder coupling constant $u_0$ is given by
 \bal
\delta\text{-correlated:}  \hspace{1cm}  \big[ u_0 \big] &= 2-d, \\
 \text{power-law:} \hspace{1cm}  \big[ u_0 \big] &= 4-d-\alpha.
 \eal
It is convenient to set $\alpha = 1$ for power-law correlated disorder, because in this case all coupling constants are marginal in $d=3$ and a systematic $\epsilon$-expansion in $\epsilon=3-d$ can be carried out. Even though the disorder coupling for $\delta$-correlated disorder is seemingly irrelevant in dimensions $d>2$, we will show below that the loop expansion is organized in terms of an effective coupling $u_{\delta} = u_0 \Lambda$, with $\Lambda$ the UV momentum cutoff. Note that this coupling has scaling dimension $\big[ u_\delta \big] = 3-d$, such that a systematic expansion in small $\epsilon=3-d$ is possible as well.

In the following we study the action $S+S_\text{dis}$ of the SDW critical point with quenched disorder after introducing dimensionless coupling constants by replacing $g \to g \mu^{\epsilon/2}$, $u_{1,2} \to u_{1,2} \mu^{\epsilon}$ and $u_0 \to u_0 \mu^{\epsilon-1}$ for the $\delta$-correlated model or $u_0 \to u_0 \mu^{\epsilon}$ for power-law correlations with $\alpha=1$, with $\mu$ an arbitrary mass scale.

\section{Results}
\label{sec3}

Here we focus on results for the $\delta$-correlated disorder model and discuss differences for power-law correlated disorder at the end of this section.

\subsection{Organization of the perturbation series in $u_0 \Lambda$}
\label{sec3a}

Due to the fact the fermionic dispersion relations in Eq.~\eqref{fermiondisp} are linearised and because quadratic terms are irrelevant in the RG sense, the diagrammatic perturbation series for $\delta$-correlated disorder turns out to be organised in powers of $u_\delta = u_0 \Lambda$, with $\Lambda$ the UV momentum cutoff. Indeed, each loop integral which only contains fermion propagators with the same dispersion relation (i.e.~fermions at the same hot-spot and/or at anti-podal hot-spots) as well as disorder vertices, exhibits a trivial linear dependence on the UV momentum cutoff $\Lambda$. This can be easily seen by shifting the loop momentum, e.g.~via $k_y \to k_y -v k_x$ for loops which only contain propagators with dispersion $\varepsilon_1(k)$, and similarly for loops with other fermion dispersions. The loop integrand is thus trivially independent of $k_x$, which leads to a linear dependence on the UV-cutoff $\Lambda$. 

Crucially, such linearly diverging loop integrals cannot involve bosonic propagators or Yukawa vertices, which mix fermions at different hot-spots with different dispersion relations. For this reason linearly diverging loops only appear in combination with a disorder interaction vertex $u_0$ and the linear $\Lambda$ dependence can be absorbed into the disorder coupling $u_\delta$ as defined above, increasing its tree-level scaling dimension to $[u_\delta] = 3-d$. Importantly, this allows us to study both the Yukawa and the disorder interaction on equal footing in an epsilon expansion around $d=3$ dimensions.

Note that sub-leading diagrams exist as well, which contain disorder vertices $u_0$, but no accompanying linear diverging loop integrals, e.g.~if these loops contain boson propagators. Such diagrams are irrelevant at low energies and thus do not appear in our analysis.

\subsection{One-loop RG flow equations for $\delta$-correlated disorder}
	
	\begin{figure}
	\centering
	\includegraphics[width=.9 \columnwidth]{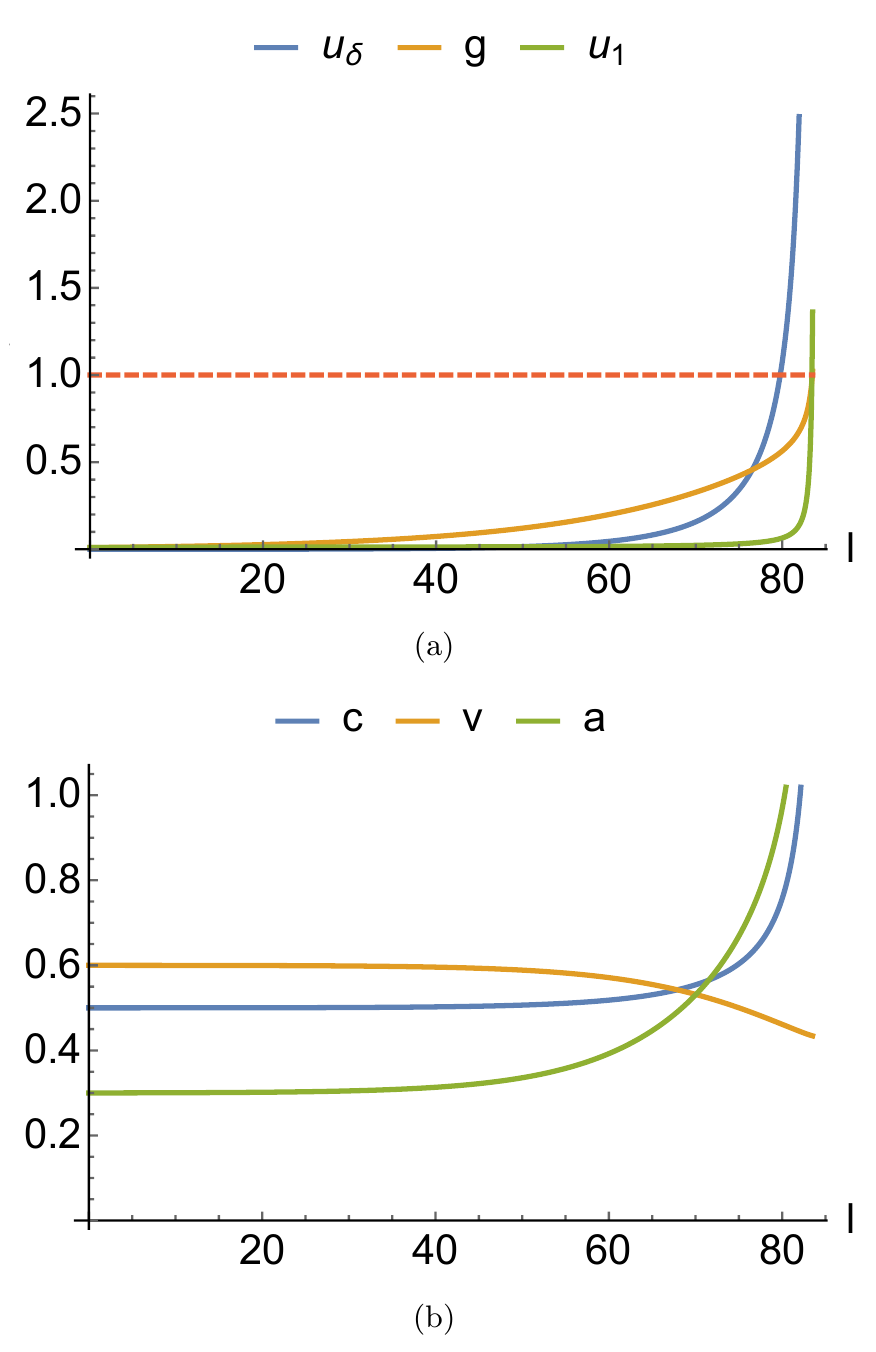}
	\caption{RG flows as function of the logarithmic energy scale $\ell=\log(\mu_0/\mu)$. In (a) we plot the flow of the dimensionless disorder coupling $u_\delta$, the Yukawa-coupling $g$ and the $\phi^4$ interaction $u_1$, whereas (b) shows plots of the Fermion-velocity $v$ and the boson velocities $c$ and $a$. Initial conditions: $(u_{\delta}, g, u_{1}, c, v, a) = $ $(0.01, 0.01, 0.0001, 0.5, 0.6, 0.3)$ and $\epsilon = 0.1$, $N_f =1$, $N_c=2$. For $N_c=2$, the two $\phi^4$ interaction terms are identical up to a factor of two and $u_2$ can be set to zero for simplicity. The dashed red line in (a) indicates where the perturbation theory formally breaks down.}
	\label{flowmodel1}
	\end{figure}

In this section we present one loop RG flow equations for all dimensionless coupling constants.
Detailed calculations of the one-loop diagrams and the renormalization procedure are presented in the appendix.
The flow equations for the velocities and the (effective) dimensionless coupling constants for $\delta$-correlated disorder potentials read	
	\begin{widetext}
	\bal
		\frac{\partial c}{\partial \ell}&= -\frac{g^2}{v} \frac{1}{8\pi^2 N_f N_c} \left\{ c N_f N_c \pi - 2v \left( N_c^2-1 \right) \left[ h_1(c,v,a)-h_2(c,v,a) \right] \right\} + \frac{c u_\delta}{\pi^2}, 
		\label{floweqsfullc}
		\\
		\frac{\partial v}{\partial \ell} &= -\frac{g^2 v}{c} \frac{N_c^2-1}{2\pi^2 N_f N_c} h_2(c,v,a), \\
		\frac{\partial a}{\partial \ell} &= -\frac{g^2}{c v a} \frac{1}{8 \pi^2 N_f N_c}\left\{ \left( a^2-1 \right)c \pi N_f N_c-2 a^2 v \left( N_c^2-1 \right) \left[ h_1(c,v,a) - h_q(c,v,a) \right] \right\} + \frac{a u_\delta}{\pi^2}, \\
		\frac{\partial g}{\partial \ell} &= \frac{\epsilon}{2} g - \frac{g^3}{c v}\frac{1}{8\pi^3 N_f N_c} \bm{\big(} c \pi^2 N_f N_c - v \left\{ h_3(c,v,a)+ \pi \left( N_c^2-1 \right) \left[ h_1(c,v,a) - h_q(c,v,a)- 2 h_2(c,v,a) \right] \right\} \bm{\big)} + \frac{g u_\delta}{2\pi^2}, \\
		\frac{\partial u_1}{\partial \ell} &= \epsilon u_1 - \frac{1}{2 \pi^2 a c^2} \left[ \left( N_c^2+7 \right) u_1^2 + \frac{2(2N_c^2-3)}{N_c} u_1 u_2 + \frac{3(N_c^2+3)}{N_c^2} u_2^2 \right] - \frac{u_1 g^2}{c v}\frac{1}{4\pi^2 N_f N_c} \big\{2\pi c N_f N_c \nn
		&\quad + v \left( N_c^2-1 \right) \left[ -3 h_1(c,v,a)+ h_q(c,v,a)+ 2 h_2(c,v,a) \right] \big\}+ \frac{3}{\pi^2} u_1 u_\delta, \\
		\frac{\partial u_2}{\partial \ell}&=  \epsilon u_2- \frac{u_2}{c^2 a \pi^2} \left( 6 u_1+\frac{N_c^2-9}{N_c}u_2 \right)- \frac{u_2 g^2}{c v}\frac{1}{4\pi^2 N_f N_c} \big\{ 2\pi c N_f N_c + v \left( N_c^2-1 \right) [ -3 h_1(c,v,a)+ h_q(c,v,a) \nn
		&\quad+ 2 h_2(c,v,a) ] \big\} + \frac{3}{\pi^2} u_2 u_\delta, \\
		\frac{\partial u_\delta}{\partial \ell} &= \epsilon u_\delta + \frac{2 u_\delta^2}{\pi^2} +\frac{u_\delta g^2}{c} \frac{N_c^2-1}{4\pi^2 N_f N_c} \left[ 2h_1(c,v,a)-h_q(c,v,a)-h_2(c,v,a) \right],
		\label{floweqsfullud}
	\eal
	\end{widetext}
where the dimensionless disorder couplings $u_\delta$ and $u_0$ (see Sec.~\ref{sec3a}) are related via $u_\delta = u_0 \Lambda/\mu$. Moreover, $\ell = \log(\mu_0/\mu)$ is a logarithmic energy scale, $c,v,a,u_i>0$ and the functions $h_i(c,v,a)$ are defined in the appendix. 
Note that these flow equations reduce to the ones found by Sur and Lee in Ref.~\cite{Sur2015} for $u_\delta=0$ and $a=1$. The boson velocity $c$ does not flow to zero in the presence of disorder, thus the arguments in Ref.~\cite{LuntsSchliefLee} that certain two-loop diagrams are important at leading order in $\epsilon$ for the clean SDW critical point, don't play a role here. 

Unfortunately these flow equations do not feature a stable fixed point at finite $u_\delta$.
Rather the RG flow is generically directed to strong coupling, with all parameters apart from the velocity $v$ diverging at low energies, where our perturbative approach eventually breaks down (see Fig.~\ref{flowmodel1}). 

	\begin{figure*}
			\centering
			\includegraphics[width=1\textwidth]{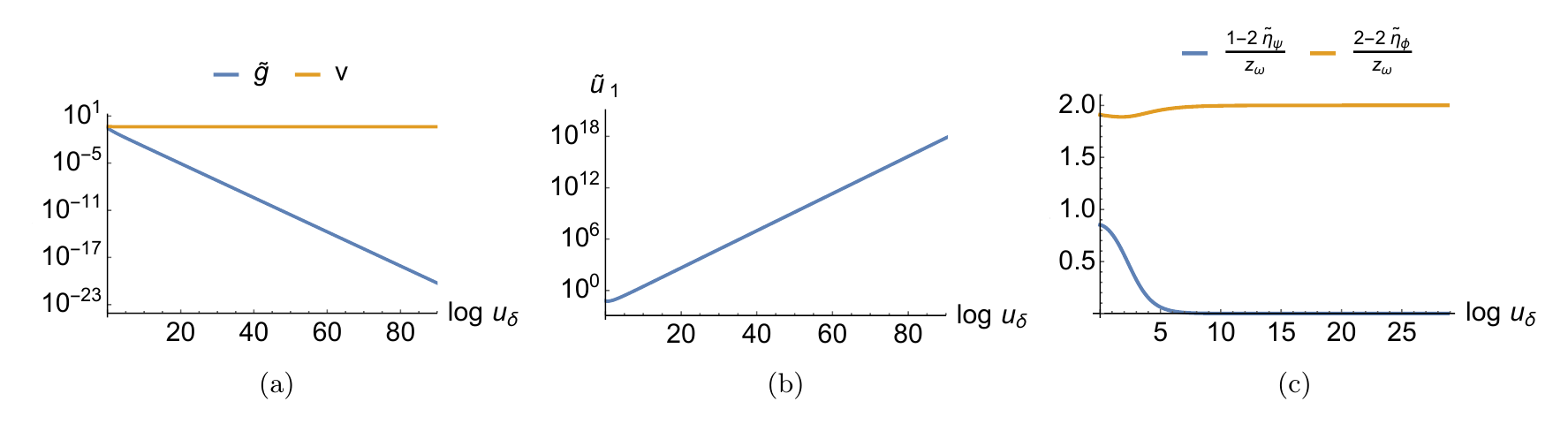}
		\caption{RG flows as function of $\log u_\delta$ (see main text for details). (a): flow of $\tilde{g} = g^2/u_\delta$ and $v$, (b) $\tilde{u}_1=u_1/u_\delta$ and (c) the frequency exponents of the fermionic and bosonic propagators. The initial conditions in (a) and (b) are taken from Fig.~\ref{flowmodel1} at $u_\delta = 1$, i.e.~$(\tilde{g}, \tilde{u}_{1}, c, v, a) = (0.3035, 0.0600, 0.7370,0.4641, 0.9347)$ for which the fermion velocity converges to $v=0.4325$. The frequency exponents in (c) are plotted for the same initial values as in Fig.~\ref{flowmodel1}, but with $\epsilon=1$. For comparison, at the clean SDW fixed point the exponents are given by $\frac{1-2\tilde{\eta}_\Psi}{z_\omega} = 1-\frac{5}{6} \epsilon$ and $\frac{2-2\tilde{\eta}_\psi}{z_\omega} = 2-\frac{5}{3}\epsilon$ at leading order in $\epsilon$.}
		\label{flowmodel1logscale}
	\end{figure*}

It is worthwhile to emphasise that our flow equations for the Yukawa coupling $g$ and the disorder coupling $u_\delta$ are formally similar to the flow equations for the $\phi^4$ interaction and the disorder coupling in the Hertz-theory of metallic SDW quantum critical points with quenched disorder, where the fermions are integrated out. This theory was analysed in a double expansion in Ref.~\cite{Kirkpatrick}. As the latter flow equations do feature novel fixed points at finite disorder and interactions, it will be elucidating to analyse the differences to our case. In order to facilitate an analytic analysis we make a simplifying assumption and set the boson velocity in the additional dimensions $a$ to its value at the clean SDW fixed point, $a=1$, where $h_q \equiv h_1$. In the hyperplane $\beta_c = 0$, where potential fixed points of the boson velocity $c$ are located, the flow equation for $u_\delta$ can then be simplified and takes the form
\beq
\frac{\partial u_\delta}{\partial \ell} = \epsilon u_\delta + \frac{u_\delta^2}{\pi^2} + \frac{u_\delta g^2}{8 \pi v} \ ,
\label{flowudeltasimple}
\eeq
whereas the flow equation for the Yukawa coupling generically reads
\beq
\frac{\partial g}{\partial \ell} = \frac{\epsilon}{2} g - \kappa\, g^3 + \frac{u_\delta g}{2\pi^2},
\eeq
where $\kappa$ depends on $c$, $v$, $N_f$ and $N_c$. As mentioned above, these flow equations are strikingly similar to the ones found in Ref.~\cite{Kirkpatrick}, which do feature a fixed point at finite $u_\delta$. The main difference in our case is that the last term in Eq.~\eqref{flowudeltasimple} comes with a plus sign, i.e.~the Yukawa coupling $g$ anti-screens and thus enhances the disorder interaction at low energies, in contrast to Ref.~\cite{Kirkpatrick}, where the disorder coupling is screened by the $\phi^4$ interaction. This anti-screening is the main reason why our theory doesn't allow for a fixed point at finite disorder and flows to strong coupling at low energies. A secondary reason is the dynamical nesting of the Fermi surface due to the flow of $v$ to smaller values, which enhances the anti-screening and also would lead to a divergence of $\kappa$ in the limit of $v \to 0$.

Even though we cannot access a potential strong disorder fixed point in our perturbative approach, we can attempt to extract some information about the behaviour of the flow towards strong coupling. Since $u_\delta$ diverges at a finite value of the logarithmic flow parameter $\ell$ and the dimensionless coupling constants are monotonic functions of $\ell$ in the regime where $u_\delta$ diverges, it is convenient to parametrise their flow in terms of $ \log u_\delta$ instead \cite{Nandkishore2017}. Corresponding flow equations can be found in Appendix~\ref{appC}.

In Fig.~\ref{flowmodel1logscale} we show plots of $g^2/u_\delta$, $v$ and $u_1/u_\delta$ as functions of $\log u_\delta$. Note that $g^2/u_\delta$ vanishes asymptotically along the flow to strong coupling, indicating that disorder dominates the Yukawa coupling at low energies. In contrast to the clean non-Fermi liquid fixed point, the fermion velocity $v$ does not flow to zero, but to a constant value instead, indicating that the hot-spots are not perfectly nested at low energies. This can be traced back to the fact that $g^2/u_\delta$ vanishes sufficiently fast along the flow to strong coupling, such that the flow of $v$ stops before it vanishes. Lastly, the $\phi^4$ interaction $u_1$ diverges even faster than the disorder interaction $u_\delta$ along the flow to strong coupling. This can be interpreted as a tendency towards local moment formation, indicating that the ground state might be in a random singlet phase \cite{BhattFisher}.

Moreover, we can analyse the scaling of the fermion and boson propagators along the flow to strong coupling. In two dimensions the fermionic and bosonic propagators scale as a function of frequency like (see Appendix \ref{appC})
	\bal
		G(\omega) &\sim \omega^{-\frac{1-2\tilde{\eta}_\Psi}{z_\omega}}, \\
		D(\omega) &\sim \omega^{-\frac{2-2\tilde{\eta}_\phi}{z_\omega}}.
	\eal
Since the dynamical critical exponent and the effective anomalous dimensions flow to the asymptotic values $z_\omega \simeq u_{\delta}/\pi^2$, $\tilde{\eta}_\Psi \sim \mathcal{O}(g^2)$, $\tilde{\eta}_\phi \simeq - u_{\delta}/\pi^2$ for large $u_{\delta/p}$, the frequency exponent of the fermionic propagator approaches zero along the flow to strong coupling, whereas the bosonic propagator keeps its bare frequency dependence $D(\omega) \sim \omega^{-2}$. An example of the flows of the exponents are shown in Fig.~\ref{flowmodel1logscale}(c). The fact that $G(\omega) \sim \text{const.}$ is reminiscent of a simple constant impurity scattering rate, which overshadows frequency dependent power-laws in the electron self-energy due to the interaction with order-parameter fluctuations. 
It is also worth mentioning that our results for the fermion propagator do not show indications of Sachdev-Ye-Kitaev (SYK) scaling (i.e.~$G(\omega) \sim \omega^{-1/2}$) as we approach strong coupling \cite{SachdevSYK}. 

Together with the flow of $g^2/u_\delta$ to zero, these results indicate that disorder effects dominate over the interaction between electrons and order parameter fluctuations at low energies. Since electron interactions seem to play a minor role here, a likely scenario for the fate of the $2d$ SDW QCP with quenched disorder is an Anderson-localized, insulating electronic ground state.

\subsection{Power-law correlated disorder}
	
Here we briefly discuss the power-law correlated disorder model with $\alpha=1$, where the disorder coupling constant $u_0$ has the same scaling dimension as the Yukawa coupling $g$, i.e.~ $[u_0]=[g]=\epsilon$. In contrast to the discussion in Sec.~\ref{sec3a}, the linear UV divergences $\sim \Lambda$  in certain loop integrals for $\delta$-correlated disorder turn into additional $\sim\log \Lambda$ divergences in the power-law correlated case, due to the $|\mathbf{p}|^{-1}$ dependence in Eq.~\eqref{Eqdisordermodels2}. We absorb these additional $\log$-divergences into the disorder coupling constant by defining $u_p = u_0 \log \Lambda$. Details of the calculations can be found in the appendices. 

The flow equations for the model with power-law correlated disorder with $\alpha=1$ have the same form as in Eqs.~\eqref{floweqsfullc}-\eqref{floweqsfullud} with the replacement $u_\delta \rightarrow \frac{u_p}{\sqrt{1+v^2}}$.
In analogy to the $\delta$-correlated case, for any positive, non-zero UV initial condition for the disorder coupling $u_p$ we always observe a flow to strong coupling. Consequently the same conclusions can be drawn as for the $\delta$-correlated case.

\section{Discussion and Conclusion}
\label{sec4}

In this work we presented a controlled, perturbative RG analysis of antiferromagnetic quantum critical points in two-dimensional metals with quenched disorder. We considered disorder models with small angle scattering only, such that cold electrons far away from the hot-spots cannot become hot by scattering off the disorder potential and thus the hot-spot model remains valid. Adopting the $\epsilon$-expansion by Sur and Lee, we derived one-loop flow equations and showed that the clean non-Fermi liquid fixed point is unstable in the presence of disorder and the theory flows to strong coupling. Extrapolating the flow to strong coupling and studying its asymptotics, we concluded that the ground state in two dimensions is potentially Anderson-localized and thus non-metallic. Since the strong coupling fixed point is not accessible in our approach, it would be interesting to see if our results can be tested using sign-problem free determinant quantum Monte-Carlo simulations \cite{Berg, Berg2018}, where disorder can be included in principle.
  
\acknowledgments

We acknowledge support from the Deutsche Forschungsgemeinschaft under Germany's Excellence Strategy-EXC-2111-390814868. J. H. acknowledges funding from the International Max Planck Research School for Quantum Science and Technology (IMPRS-QST).

\widetext
\appendix

\begingroup
\allowdisplaybreaks

\section{One-loop corrections without disorder}
\label{appA}

In this appendix, we derive the divergent parts of the self-energies and vertex corrections coming from the Yukawa and the $\phi^4$ interaction at one-loop order. Compared to the results in Ref.~\cite{Sur2015}, the $\epsilon$-poles are slightly modified due to the additional boson velocity $a$, which only plays a role as soon as disorder is taken into account. Diagrams are shown in Fig.~\ref{diagramswithoutdisorder}.

\subsection{Fermion self-energy}

The one-loop fermion self-energy due to the Yukawa interaction is given by the integral
	\bal
		\Sigma_{Y,n}^{(ab)}(q) = &-2 i g^2\mu^\epsilon \frac{N_c^2-1}{N_f N_c} \int_k D^{\alpha \beta}(k) \gamma^\alpha_{(ac)} \sigma_x G^{(cd)}_{\bar{n}}(q-k) \sigma_x \gamma^{\beta}_{(db)}
	\eal
and has the same causality structure as the inverse fermion propagator
	\bal
		\Sigma_{Y,n}^{(ab)} = \bmat \Sigma_{Y,n}^{R} & \Sigma_{Y,n}^{K} \\ 0 & \Sigma_{Y,n}^{A} \emat.
	\eal
Indeed, the $\Sigma_{Y,n}^{(21)}$-component 
	\bal
		\Sigma_{Y,n}^{(21)} \sim \int_\omega \left[ D^R(k) G^A_{\bar{n}}(q-k) + D^A(k) G^R_{\bar{n}}(q-k) \right]
	\eal
vanishes by integrating over the frequency since all poles of the propagators lie in the same half plane.
	\begin{figure*}
			\centering
			\includegraphics[width=1\textwidth]{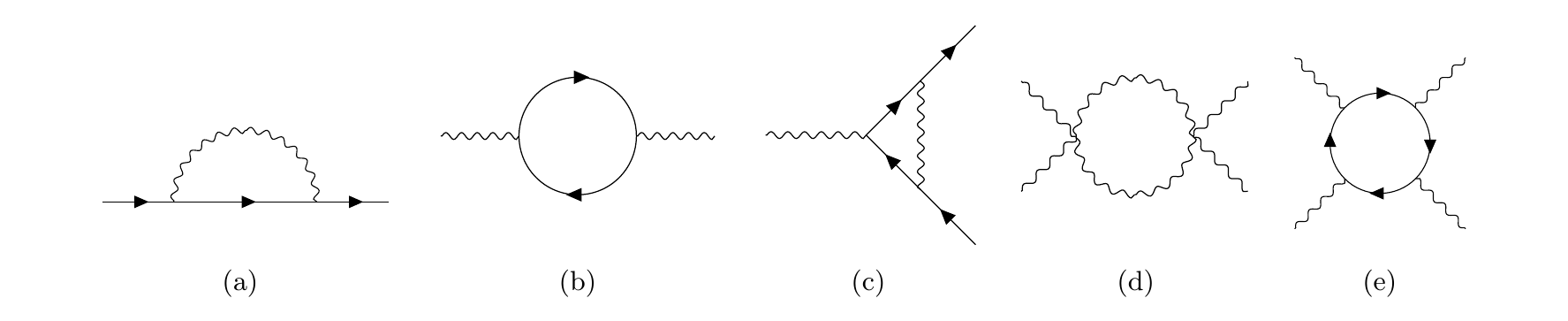}
		\caption{One-loop contributions from Yukawa and $\phi^4$ interactions to (a) fermionic self-energy, (b) bosonic self-energy, (c) Yukawa vertex and (d), (e) $\phi^4$ vertices. Only diagram (e) does not lead to a pole in $\epsilon=3-d$. Solid (wiggly) lines represent fermionic (bosonic) propagators. All propagators are matrices in the Keldysh-index space.}
		\label{diagramswithoutdisorder}
	\end{figure*}
For the remaining components it is sufficient to calculate the retarded self-energy
	\bal
		&\Sigma_{Y,n}^{R}(q)=-2 i g^2\mu^\epsilon \frac{N_c^2-1}{N_f N_c} \int_k \sigma_x \left[ D^K(k) G^{R}_{\bar{n}}(q-k)+D^R(k) G^{K}_{\bar{n}}(q-k) \right] \sigma_x,
	\eal
since $\Sigma^A_{Y,n}$ and $\Sigma^K_{Y,n}$ can be obtained via the relations $\sigma_y \Sigma^A_{Y,n}(q)=\left[ \sigma_y \Sigma^R_{Y,n}(q) \right]^\dagger$ and $\Sigma^K_{Y,n}(q)= \sgn{q_0} \left[ \Sigma^R_{Y,n}(q) - \Sigma^A_{Y,n}(q) \right]$, respectively. 

Writing the Keldysh propagators in terms of the retarded and advanced propagators, the retarded fermion self-energy can be brought into the form
	\bal
		\Sigma_{Y,n}^{R}(q)=-\frac{2}{\pi} i g^2\mu^\epsilon \frac{N_c^2-1}{N_f N_c} \int_{\tk,\k} \sigma_x \bigg[ &\int_{q_0}^\infty d\omega D^R(k) G^{A}_{\bar{n}}(q-k) - \int_0^\infty d\omega D^A(k) G^{R}_{\bar{n}}(q-k) \nn
		&+\int_0^{q_0} d\omega D^R(k) G^{R}_{\bar{n}}(q-k) \bigg] \sigma_x.
	\eal
A pole in $\epsilon=3-d$ can be only obtained by sending one integration bound to infinity, whereas finite integration bounds cannot contribute to a divergence of the integral. Thus, the last term can be neglected safely and the remaining two terms may be written as
	\bal
		&\Sigma_{Y,n}^{R}(q) = -\frac{2}{\pi} i g^2\mu^\epsilon \frac{N_c^2-1}{N_f N_c} \int_{\tk,\k} \int_{0}^\infty d\omega\ \sigma_x \bigg[ D^R(k) G^{A}_{\bar{n}}(q-k) - D^A(k) G^{R}_{\bar{n}}(q-k) \bigg] \sigma_x \nn
		&= -\frac{2}{\pi} i g^2\mu^\epsilon \frac{N_c^2-1}{N_f N_c} \int_{\tk,\k} \int_{0}^\infty d\omega\ \tn{Im}\left\{ \frac{1}{\left( \omega+i\delta \right)^2-a^2\tk^2-c^2\k^2} \frac{i\sigma_y \left( \omega-q_0+i\delta \right) +\boldsymbol{\sigma}_{\mathbf{z}} (\tk-\tq)-\sigma_x \varepsilon_{\bar{n}}(\k-\q)}{\left( \omega-q_0+i\delta \right)^2-(\tk-\tq)^2-\varepsilon^2_{\bar{n}}(\k-\q)} \right\}.
	\eal
Introducing a Feynman parameter and completing the squares in the denominator, we find
	\bal
		&\Sigma_{Y,n}^{R}(q) = - i \frac{g^2\mu^\epsilon}{c} \frac{N_c^2-1}{N_f N_c} \frac{1}{2^{d-2}\pi^{\frac{d}{2}+1} \eulergamma{\frac{d}{2}}} \int_0^1 dx \frac{\left[ x+(1-x)a^2 \right]^{\frac{2-d}{2}}}{\sqrt{1-x} \sqrt{c^2+x(1+v^2-c^2)}} \int_0^\infty dr\ r^{d-1} \int_{-x q_0}^\infty d\omega \nn
	&\tn{Im}\left\{ \frac{i \sigma_y \left[\omega+i\delta -(1-x)q_0 \right] - \boldsymbol{\sigma}_{\mathbf{z}} \frac{(1-x)a^2}{x+(1-x)a^2}\tq + \sigma_x \frac{(1-x)c^2}{c^2+x(1+v^2-c^2)} \varepsilon_{\bar{n}}(\q)}{\left[ (\omega+i\delta)^2-r^2- \bar{q} \right]^2} \right\},
	\eal
where $\bar{q}= \bar{q}(x,c,v,a,q)$ is a real function of the Feynman parameter, the velocities, the external frequency and the external momenta. Evaluating the remaining integrals and expanding in small $\epsilon$ yields the pole
	\bal
		\Sigma_{Y,n}^{R}(q) = - \frac{g^2}{4\pi^2c\epsilon} \frac{N_c^2-1}{N_f N_c} \Big[ \sigma_y h_1(c,v,a) q_0 - i \boldsymbol{\sigma}_{\mathbf{z}} h_q(c,v,a) \tq +i \sigma_x h_2(c,v,a) \varepsilon_{\bar{n}}(\q) \Big] 
	\eal
with
	\bal
		h_1(c,v,a) &= \int_0^1 dx \sqrt{\frac{1-x}{\left[c^2+x(1+v^2-c^2)\right] \left[x+(1-x)a^2\right]}},\\
		h_q(c,v,a) &= a^2 \int_0^1 dx \sqrt{\frac{1-x}{\left[ c^2+x(1+v^2-c^2)\right] \left[ x+(1-x)a^2 \right]^3}}, \\
		h_2(c,v,a) &= c^2 \int_0^1 dx \sqrt{\frac{1-x}{\left[ c^2+x(1+v^2-c^2) \right]^3 \left[ x+(1-x)a^2 \right]}}.
	\eal
Correspondingly, the advanced component of the self-energy takes the same form, such that there is no $\epsilon$-pole for the Keldysh self-energy.

\subsection{Boson self-energy}

The boson self-energy 
	\bal
		\Pi^{\alpha \beta}(q) &= 2ig^2 \mu^\epsilon \sum_n \int_k \tn{Tr}\left\{ \gamma^\alpha_{(ab)} \sigma_x G^{(bc)}_{\bar{n}}(k) \gamma^\beta_{(cd)} \sigma_x G^{(da)}_n(k-p) \right\}
	\eal
is independent of the velocity $a$ since the integral only contains fermionic propagators. Thus, the divergent parts of the retarded and advanced self-energies can be derived to take the form
	\bal
		\Pi^{qc}(q)=\Pi^{R}(q)= \Pi^{A}(-q) = -\frac{g^2}{2\pi v \epsilon}\left( q_0^2- \tq^2 \right),
	\eal
whereas the Keldysh component $\Pi^{K}(q)=\sgn{q_0} \left[ \Pi^{R}(q)-\Pi^{A}(q) \right]$ remains finite again.

\subsection{Yukawa vertex correction}

The counter term to the Yukawa vertex coming from corrections due to the Yukawa vertex itself takes the form
	\bal
		S_{Y,\tn{count}} &=- i \frac{g \mu^{\frac{\epsilon}{2}}}{\sqrt{N_f}} \sum_{n,j} \sum_{\sigma_1,\sigma_2} \int_{p,q} \vec{\phi}^\alpha(q) \vec{\tau}_{\sigma_1 \sigma_2} \Psib^{(a)}_{n,j,\sigma_1}(p+q) V^{\alpha,(ab)}_{Y,n} \Psi^{(b)}_{\bar{n},j,\sigma_2}(p) \label{Yukawacounterterm},
	\eal
where
	\bal
		V^{\alpha,(ab)}_{Y,n} &= -2i \frac{g^2 \mu^\epsilon}{N_f N_c} \int_k \sigma_x \gamma^{\beta_1}_{(ac)} G^{(cd)}_{\bar{n}} (p+q-k) \sigma_x \gamma^{\alpha}_{(de)}D^{\beta_1 \beta_2}(k) G^{(ef)}_n(p-k) \sigma_x \gamma^{\beta_2}_{(fb)}
	\eal
has eight components corresponding to the different Keldysh indices. In fact, $V^{c,(12)}_{n}$, $V^{c,(21)}_{n}$, $V^{q,(11)}_{n}$ and $V^{q,(22)}_{n}$ turn out to be zero, whereas the remaining four components lead to the same integral expression. Hence we can write $V^{\alpha,(ab)}_{Y,n} = \gamma^\alpha_{(ab)} V_{Y,n}$ with
	\bal
		&V_{Y,n}= -4 i \frac{g^2 \mu^\epsilon}{N_f N_c} \int_{\tk,\k} \int_0^\infty \frac{d\omega}{2\pi}\Big[ D^R(k) \sigma_x G^A_{\bar{n}}(p+q-k) \sigma_x G^A_n(p-k) \sigma_x - D^A(k) \sigma_x G^R_{\bar{n}}(p+q-k) \sigma_x G^R_n(p-k) \sigma_x \Big] \nn
		&=-4 \frac{g^2 \mu^\epsilon}{N_f N_c} \sigma_x \nn 
		&\tn{Im}\left\{ \int_{\tk,\k} \int_0^\infty \frac{d\omega}{2\pi} \frac{(\omega+i\delta)^2-(\tk-\tp)^2+\varepsilon_{\bar{n}}(\k) \varepsilon_n(\k)}{\left[ (\omega+i\delta)^2 - (\tk-\tp)^2-\varepsilon_{\bar{n}}^2(\k) \right] \left[ (\omega+i\delta)^2 - (\tk-\tp)^2-\varepsilon_{n}^2(\k) \right] \left[ (\omega+i\delta)^2-a^2\tk^2-c^2\k^2 \right]} \right\},
	\eal	
where we set all external frequencies and momenta to zero except for $\tp$. Changing integration variables to $\varepsilon_{\bar{n}}(\k) = \sqrt{2v}R \sin\theta$, $\varepsilon_{n}(\k) = \sqrt{2v}R \cos\theta$ and introducing two Feynman parameters, the divergent part of the integral can be extracted from
\endgroup
	\bal
		&V_{Y,n}= -\frac{1}{2^{d-3}\pi^{\frac{d}{2}+2} \eulergamma{\frac{d-2}{2}}} \frac{g^2 \mu^\epsilon}{N_f N_c} \sigma_x \nn
		&\tn{Im}\left\{ \int_0^1dx\ dy\ (1-y) \int_0^{2\pi} d\theta \int_0^\infty dr dR\ r^{d-3}\ R\ \frac{(\omega+i\delta)^2 - r^2+v R^2 \sin(2\theta)}{\left\{ (\omega+i\delta)^2-[1+x(1-x)a^2]r^2- c R^2\zeta(c,v,x,y,\theta) - \bar{p}^2 \right\}^3} \right\}
	\eal
\begingroup
\allowdisplaybreaks
with $\zeta(c,v,x,y,\theta)=\frac{2v}{c}(1-y)\left[ x \cos^2\theta+(1-x)\sin^2\theta \right]+ \frac{c}{2v} y \left( \cos\theta+\sin\theta \right)^2+\frac{cv}{2}y \left( \cos\theta-\sin\theta \right)^2$ and $\bar{p}^2=\frac{y(1-y)a^2}{1+(a^2-1)y}\tp^2$. Evaluating the remaining integrals and expanding in small $\epsilon$, we obtain the pole
	\bal
		V_{Y,n}=- \frac{g^2}{8\pi^3 c N_f N_c \epsilon} h_3(c,v,a) \sigma_x
	\eal
with
	\bal
		h_3(c,v,a)= \frac{1}{2} \int_0^1 dx\ dy\ \frac{1-y}{\sqrt{1+(a^2-1)y}}\int_0^{2\pi}d\theta \frac{1}{\zeta(c,v,x,y,\theta)}\left[ \frac{2+(a^2-1)y}{1+(a^2-1)y}-\frac{2v \sin(2\theta)}{c \zeta(c,v,x,y,\theta)} \right].
	\eal

\subsection{$\vec{\phi}^4$ vertex correction}

The one-loop diagrams leading to counter terms for the $\vec{\phi}^4$ vertices contain only bosonic propagators. For a general second boson velocity $a$, we therefore obtain an additional factor of $a^{-1}$ in $d=3$ by rescaling the integration variable $\tk \rightarrow a^{-1} \tk$ in the boson propagators. Consequently, the counter terms to the $\vec{\phi}^4$ vertices read
	\bal
		S_{\phi^4,\tn{count}} = &-\frac{u_1 \mu^\epsilon}{4} V_{\phi^4,u_1} \int_{p_1,p_2,p_3}  \tn{Tr} \left\{ \Phi^c (p_1+p_3) \Phi^c(p_2-p_3) + \Phi^q (p_1+p_3) \Phi^q(p_2-p_3) \right\} \tn{Tr} \left\{ \Phi^c (-p_1) \Phi^q(-p_2) \right\} \nn
	&-\frac{u_2 \mu^\epsilon}{4} V_{\phi^4,u_2} \int_{p_1,p_2,p_3}  \tn{Tr} \left\{ \left[ \Phi^c (p_1+p_3) \Phi^c(p_2-p_3) + \Phi^q (p_1+p_3) \Phi^q(p_2-p_3)\right] \Phi^c (-p_1) \Phi^q(-p_2) \right\}
	\eal
with
	\bal
		V_{\phi^4,u_1} &= \frac{1}{2\pi^2 c^2 a} \Bigg[ \frac{u_1}{\epsilon} (N_c^2+7)+\frac{u_2}{\epsilon} \frac{2(2N_c^2-3)}{N_c} + \frac{u_2^2}{u_1 \epsilon} \frac{3(N_c^2+3)}{N_c^2} \Bigg],\\
		V_{\phi^4,u_2} &= \frac{1}{\pi^2 c^2 a} \left[ \frac{6 u_1}{\epsilon} + \frac{u_2}{\epsilon} \frac{N_c^2-9}{N_c} \right].
	\eal

\section{One-loop corrections including disorder}
\label{appB}

\subsection{Fermion self-energy}

	\begin{figure*}
			\centering
			\includegraphics[width=1\textwidth]{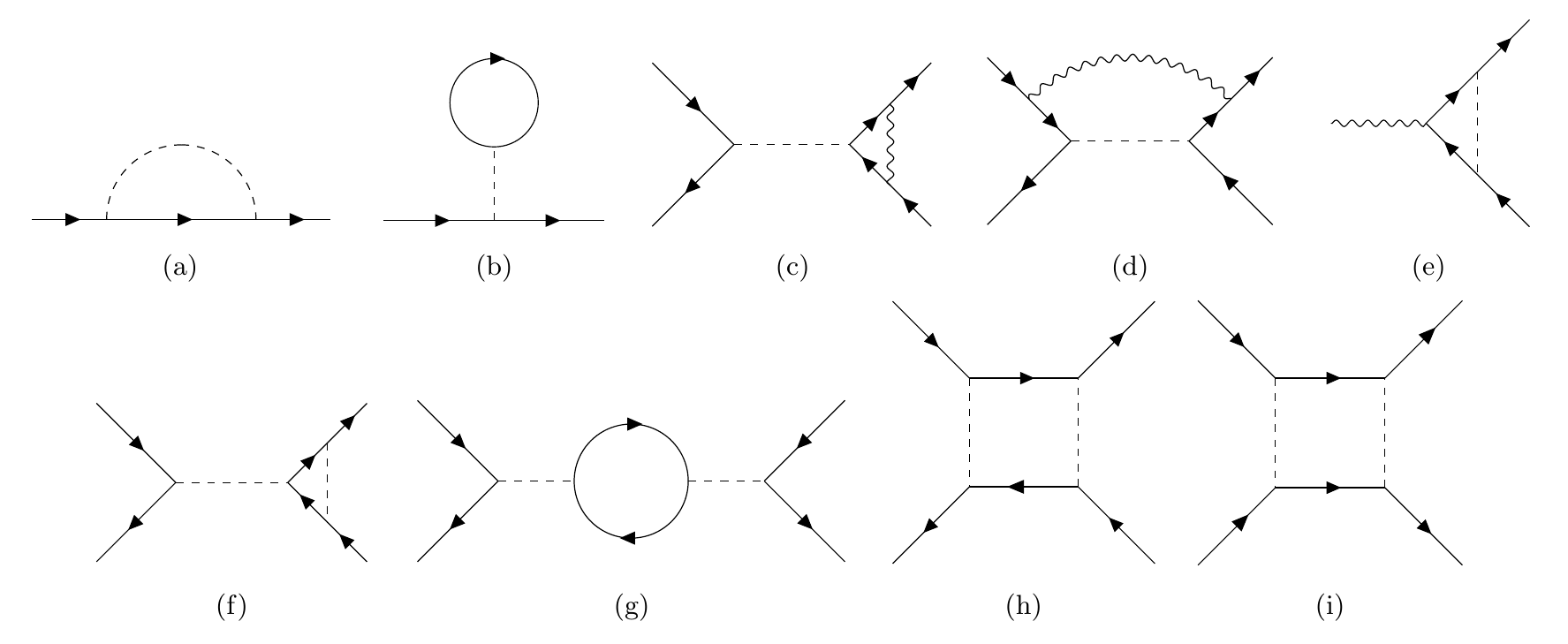}
		\caption{One-loop contributions including disorder to (a, b) fermionic self-energy, (c, d) disorder vertex coming from Yukawa interaction, (e) Yukawa vertex coming from disorder interaction and (f)-(i) disorder vertex from disorder interaction. Dashed lines represent factors of $u(\vec{k})$.}
		\label{diagramswithdisorder}
	\end{figure*}
The fermion self-energy obtained from the disorder vertex (Fig.~\ref{diagramswithdisorder}(a)) is given by
	\bal
		\Sigma^{(ab)}_{\tn{dis},n}(q)= 2 \int_{\vec{k}} \sigma_y G^{(ab)}_n(\vec{k},q_0) \sigma_y u(\vec{k}-\vec{q}),
	\eal
where $u(\vec{k}-\vec{q})=u_0 \mu^{\epsilon-1}$ for $\delta$-correlated disorder and $u(\vec{k}-\vec{q})=u_0 \mu^\epsilon \av{\k-\q}^{-1}$ for the power-law correlated case with $\alpha = 1$. The second contribution (Fig.~\ref{diagramswithdisorder}(b)) only leads to an unimportant renormalization of the chemical potential and will be omitted. 

\subsubsection{$\delta$-correlated disorder}

Taking the retarded/advanced component of the self-energy, the matrix product simplifies to
	\bal
		\Sigma^{R/A}_{\tn{dis},n}(q) = 2 u_0 \mu^{\epsilon-1} \int_{\vec{k}} \frac{\sigma_y(q_0\pm i\delta)+i\boldsymbol{\sigma}_{\mathbf{z}} \tk+i\sigma_x\varepsilon_n(\k)}{(q_0\pm i\delta)^2-\tk^2-\varepsilon_n^2(\k)}.
	\eal
Since the integrand only depends on a single hot-spot via $\varepsilon_n(\k)$, it is always possible to shift out one of the integration variables $k_x$ or $k_y$. Taking e.g. $n=1$ with $\varepsilon_1(k)=v k_x+k_y$ and shifting $ k_y \rightarrow k_y-v k_x$ leads to
	\bal
		\Sigma^{R/A}_{\tn{dis},n}(q) &= 2 u_0 \mu^{\epsilon-1} \frac{\Lambda}{\pi} \int_{\tk,k_y} \frac{\sigma_y(q_0\pm i\delta)+i\boldsymbol{\sigma}_{\mathbf{z}} \tk+i\sigma_x k_y}{(q_0\pm i\delta)^2-\tk^2-k_y^2},
	\eal
with $\Lambda$ being the cutoff for the $k_x$-integral. Evaluating the remaining integrals and expanding in small $\epsilon=3-d$ we obtain  
	\bal
		\Sigma^{R/A}_{\tn{dis},n}(q)= -\frac{u_{\delta}}{\pi^2\epsilon} \sigma_y q_0, \label{deltadisorderfermionselfenergy}
	\eal
where we defined the effective coupling $u_{\delta} = u_0 \Lambda/\mu$. Furthermore we dropped the dependence on $\delta=0^+$ since it is only needed to make the integrals convergent. Note that the calculation is in fact independent of the hot-spot index, such that \eqref{deltadisorderfermionselfenergy} holds for any $n$.

\subsubsection{Power-law correlated disorder}

For the power-law correlated case, the retarded/advanced fermion self-energy can be calculated from
	\bal
		\Sigma^{R/A}_{\tn{dis},n}(q) &= 2 u_0 \mu^{\epsilon} \int_{\vec{k}} \frac{1}{\sqrt{k_x^2+k_y^2}} \frac{\sigma_y(q_0\pm i\delta)+i\boldsymbol{\sigma}_{\mathbf{z}} \tk+i\sigma_x\varepsilon_n(\k+\q)}{(q_0\pm i\delta)^2-\tk^2-\varepsilon_n^2(\k+\q)}.
	\eal
Taking again $n=1$ as an example, the above integral can be brought into the form
	\bal
		&\Sigma^{R/A}_{\tn{dis},n}(q) = 2 u_0 \mu^{\epsilon} \int_{\vec{k}} \frac{1}{\sqrt{k_x^2+k_y^2}} \frac{\sigma_y(q_0\pm i\delta)+i\boldsymbol{\sigma}_{\mathbf{z}} \tk+i\sigma_x \left[ \sqrt{1+v^2} k_y+\varepsilon_1(\q)\right]}{(q_0\pm i\delta)^2-\tk^2-\left[ \sqrt{1+v^2} k_y+\varepsilon_1(\q)\right]^2},
	\eal
where the linear divergence of the $k_x$-integral in the $\delta$-correlated case is now replaced by a logarithmic divergence. Evaluating the remaining integrals thus leads to
	\bal
		\Sigma^{R/A}_{\tn{dis},n}(q)= -\frac{u_{p}}{\sqrt{1+v^2}\pi^2\epsilon} \sigma_y q_0
	\eal
with the effective interaction $u_p= u_0 \log\Lambda$. Again, the calculation is independent of the specific choice of $n$.

\subsection{Disorder vertex correction coming from disorder vertex}

The counter term for the disorder vertex coming from diagram Fig.~\ref{diagramswithdisorder}(f) reads 
	\bal
		S_{\tn{dis},\tn{count}}= i \sum_{n_1,n_2}\sum_{\sigma_1,\sigma_2} \sum_{j_1,j_2} \int_{\overset{\vec{k}_1,\vec{k}_2,\vec{k}_3}{\omega_1,\omega_2}} &\left[\Psib^{(a)}_{n_1,j_1,\sigma_1}(\vec{k}_1+\vec{k}_3,\omega_1) \sigma_y \Psi^{(a)}_{n_1,j_1,\sigma_1}(\vec{k}_1,\omega_1) \right] u(\vec{k}_3) \nn
		&\left[\Psib^{(b)}_{n_2,j_2,\sigma_2}(\vec{k}_2-\vec{k}_3,\omega_2) U^{(bc)}_{\tn{dis},n} \Psi^{(c)}_{n_2,j_2,\sigma_2}(\vec{k}_2,\omega_2) \right] \label{disordercounterterm}
	\eal
with
	\bal
		U^{(bc)}_{\tn{dis},n} = - 4 \int_{\vec{p}} \sigma_y G^{(bd)}_n(\vec{p}+\vec{k}_2-\vec{k}_3,\omega_2) \sigma_y G^{(dc)}_n(\vec{p}+\vec{k}_2,\omega_2) \sigma_y u(\vec{p}).
	\eal
Diagram Fig.~\ref{diagramswithdisorder}(g) trivially vanishes by integrating over the  internal frequency of the fermion loop since only combinations $G^{(ab)}G^{(ba)} \sim G^R G^R, G^A G^A$ contribute. The other two diagrams don't lead to a renormalization of the disorder vertex as well, which will be explained in the following subsections.

In principle, we could set the external frequency to zero implying that the exact nature of the propagators (i.e. retarded or advanced) does not matter for the calculation of the poles, s.t. any expression containing a Keldysh propagator vanishes immediately. Thus, the off-diagonals of the vertex correction $U^{\tn{dis},(bc)}_n$ are zero, whereas the diagonals are identical and non-zero. Nevertheless, we keep the external frequency finite in the following to regulate the momentum integrals, but the final expression of the poles will be independent of the frequency of course.

\subsubsection{$\delta$-correlated disorder}

Setting the external momenta to zero and $u(\vec{p})=u_0 \mu^{\epsilon-1}$, the divergent part of the diagonal entries of the vertex correction can be extracted from
	\bal
		U^{(11)}_{\tn{dis},n} = - 4 u_0 \mu^{\epsilon-1} \sigma_y \int_{\vec{p}} \frac{(\omega_2+i\delta)^2+\tp^2+\varepsilon_n^2(\p)}{\left[ (\omega_2+i\delta)^2 - \tp^2-\varepsilon_n^2(\p) \right]^2}.
	\eal
As was the case for the fermion self-energy, we can shift out one of the momenta in $\varepsilon_n(\k)$, leaving us with a linearly diverging integral, s.t. we find
	\bal
		U^{(11)}_{\tn{dis},n} &= - 4 u_0 \mu^{\epsilon-1} \sigma_y \frac{\Lambda}{\pi} \int_{\tp,p_y} \frac{(\omega_2+i\delta)^2+\tp^2+p_y^2}{\left[ (\omega_2+i\delta)^2 - \tp^2-p_y^2 \right]^2} =- \frac{2 u_{\delta}}{\pi^2 \epsilon} \sigma_y = U^{(22)}_{\tn{dis},n}.
	\eal

The hot-spot indices of the internal propagators in the two diagrams Fig.~\ref{diagramswithdisorder}(h) and Fig.~\ref{diagramswithdisorder}(i) don't have to be the same in principal. If they have the same index, we indeed find $\epsilon$-poles, but the contributions from the two diagrams exactly cancel. If they have different indices, the momentum integrals turn out to be divergent for any $d\ge 2$ since we cannot shift out one of the integration variables. In $d=3$ the divergence is linear in the cutoff $\Lambda$ (without an $\epsilon$-pole) and therefore subleading to terms $\sim \frac{\Lambda}{\epsilon} \sim \Lambda \log\Lambda$, such that their contribution can be neglected. 

\subsubsection{Power-law correlated disorder}

In the power-law correlated case, the linear divergence is again replaced by a logarithmic divergence, leading to the pole
	\bal
		U^{(11)}_{\tn{dis},n} &=U^{(22)}_{\tn{dis},n}= - \frac{2 u_{p}}{\sqrt{1+v^2}\pi^2 \epsilon} \sigma_y.
	\eal

The two diagrams Fig.~\ref{diagramswithdisorder}(h) and Fig.~\ref{diagramswithdisorder}(i) turn out to be finite for any $d\le 3$.

\subsection{Disorder vertex correction coming from Yukawa vertex}

Similarly to \eqref{disordercounterterm}, we need to add a counter term to the disorder vertex coming from corrections via the Yukawa interaction (Fig.~\ref{diagramswithdisorder}(c)), which can be calculated via
	\bal
		U^{(bc)}_{Y,n} = 4ig^2 \mu^\epsilon \frac{N_c^2-1}{N_c N_f} \int_{\omega,\vec{p}} \gamma^\alpha_{(bd)} \sigma_x G^{(de)}_{\bar{n}}(\vec{k}_2-\vec{k}_3-\vec{p},\omega) \sigma_y D^{\alpha \beta}(\vec{k},\omega_2-\omega) G^{(ef)}_{\bar{n}}(\vec{k_2}-\vec{p},\omega) \sigma_x \gamma^\beta_{(fc)}
	\eal
and is identical for both $\delta$-correlated and power-law correlated disorder. In principle three additonal diagrams (Fig.~\ref{diagramswithdisorder}(d) plus two permutations of the direction of the fermions) could possibly renormalize the disorder vertex, but the integrals are convergent in $d=3$ due to a missing internal frequency integral. 

Again, the off-diagonal contributions vanish, whereas the diagonal contributions are equal and non-zero. Simplifying the sum over Keldysh indices and setting all external momenta and frequencies to zero except for $\tk_2$, the integral reduces to
	\bal
		U^{(11)}_{Y,n}= 8 g^2 \mu^\epsilon \frac{N_c^2-1}{N_c N_f} \sigma_y \tn{Im}\left\{ \int_{\vec{p}} \int_0^\infty \frac{d\omega}{2\pi} \frac{(\omega+i\delta)^2+(\tp-\tk_2)^2+\varepsilon_{\bar{n}}^2(\p)}{\left[ (\omega+i\delta)^2-(\tp-\tk_2)^2 - \varepsilon_{\bar{n}}^2(\p) \right]^2 \Big[(\omega+i\delta)^2-a^2\tp^2-c^2\p^2 \Big]} \right\}.
	\eal
Introducing a Feynman parameter and completing the squares in the resulting numerator, the divergent part of the integral can be extracted from
	\bal
		&U^{(11)}_{Y,n}= 8 g^2 \mu^\epsilon \frac{N_c^2-1}{N_c N_f} \sigma_y \nn
		&\tn{Im}\left\{2 \int_0^1dx\ x \int_{\vec{p}} \int_0^\infty \frac{d\omega}{2\pi} \frac{(\omega+i\delta)^2+\tp^2+v^2 p_x^2 + \frac{c^4(1-x)^2}{[(1-x)c^2+xv^2]^2}p_y^2}{\left\{ (\omega+i\delta)^2 -(x+(1-x)a^2)\tp^2-[(1-x)c^2+xv^2]p_x^2-\frac{c^2(1-x)[c^2+x(1+v^2-c^2)]}{(1-x)c^2+xv^2}p_y^2 -\bar{k}_2^2 \right\}^3} \right\}
	\eal
with $\bar{k}_2^2 = \frac{x(1-x)a^2}{x+(1-x)a^2}\tk_2^2$ and evaluates to
	\bal
		&U^{(11)}_{Y,n}= 8 g^2 \mu^\epsilon \frac{N_c^2-1}{N_c N_f} \sigma_y \nn
		&\tn{Im}\left\{-\frac{i}{32\pi^2c \epsilon} \int_0^1 dx \frac{x}{\sqrt{1-x}} \frac{1}{\sqrt{c^2+x(1+v^2-c^2)} \sqrt{x+(1-x)a^2}} \left( -1+\frac{1}{x+(1-x)a^2} +\frac{1+v^2}{c^2+x(1+v^2-c^2)} \right) \right\}.
	\eal
The Feynman parameter integral turns out to be numerically the same as $2h_1(c,v,a)$, such that we obtain 
	\bal
		U^{(11)}_{Y,n} &= U^{(22)}_{Y,n} = 8 g^2 \frac{N_c^2-1}{N_c N_f} \tn{Im} \left\{ -\frac{i}{32\pi^2c \epsilon} 2h_1(c,v,a) \right\} = - \frac{g^2}{2\pi^2 c \epsilon} \frac{N_c^2-1}{N_c N_f} h_1(c,v,a).
	\eal
Note that the $\epsilon$-pole is again independent of the hot-spot index $n$.

\subsection{Yukawa vertex correction coming from disorder vertex}

We need to add a similar counter term as in \eqref{Yukawacounterterm} to cancel possible corrections to the Yukawa vertex coming from the disorder interaction. The divergent part of the one-loop contribution (Fig.~\ref{diagramswithdisorder}(e)) can be calculated from the expression
	\bal
		V^{\alpha,(ab)}_{\tn{dis},n} = -2 \int_{\vec{k}} \sigma_y G^{(ac)}_n(\vec{k}+\vec{p}+\vec{q},p_0+q_0) \gamma^{\alpha}_{(cd)} \sigma_x G^{(db)}_{\bar{n}}(\vec{k}+\vec{p},p_0)\ u(\vec{k}).
	\eal
Again, this vertex correction has the Keldysh index structure of the original Yukawa vertex $V^{\alpha,(ab)}_{\tn{dis},n} = \gamma^\alpha_{(ab)} V_{\tn{dis},n}$ with
	\bal
		V_{\tn{dis},n} = -2 \sigma_x \int_{\vec{k}} \frac{\varepsilon_{n}(\k) \varepsilon_{\bar{n}}(\k)-\tk^2}{\left[ (p_0+i\delta)^2 - \tk^2 -\varepsilon_n^2(\k) \right] \left[ (p_0+i\delta)^2 - \tk^2 -\varepsilon_{\bar{n}}^2(\k) \right]} u(\vec{k}), \label{yukawavertexcorrectionfromdis}
	\eal
where we set all external frequencies and momenta to zero except for $p_0$. The final result will be independent of the hot-spot index $n$ again, s.t. we choose $n=1$ in the following for simplicity.

\subsubsection{$\delta$-correlated case}

For a general finite fermion velocity $v$, the momentum integral is again divergent for dimensions $d\ge2$, but only with a linear $\Lambda$ divergence without an $\epsilon$-pole in d=3. Only for $v=0$, the dependence on one of the integration variables drops out and we obtain
	\bal
		\lim_{v\rightarrow 0}V_{\tn{dis},n} = \frac{u_\delta}{\pi^2 \epsilon} \sigma_x.
	\eal
Since the limit $v\rightarrow 0$ can be reached only asymptotically during the RG flow, we will neglect the above contribution in our calculations.

\subsubsection{Power-law correlated case}

In the limit of vanishing fermion velocity, the integral can be computed straightforwardly
	\bal
		V_{\tn{dis},1} &= 2 u_0 \mu^\epsilon \sigma_x \int_{\vec{k}} \frac{\tk^2+k_y^2}{\left[ (p_0+i\delta)^2 - \tk^2 -k_y^2 \right]^2} \frac{1}{\sqrt{k_x^2+k_y^2}} \nn
		&= \frac{u_p}{\pi^2\epsilon} \sigma_x. \label{yukawavertexzerov}
	\eal

For general finite $v$, the integrand \eqref{yukawavertexcorrectionfromdis} is rewritten by introducing two Feynman parameters
	\bal
		V_{\tn{dis},1} &= \frac{3}{2} u_0 \mu^\epsilon \sigma_x \int_0^1 dx dy \frac{y}{\sqrt{1-y}} \int_{\vec{k}} \frac{\tk^2+k_y^2-v^2 \left[ 1-(1-2x)^2y^2 \right] k_x^2}{\bm{\big[}  y \tk^2 + k_y^2+\bm{(}  1-y \left\{1-v^2 \left[ 1-(1-2x)^2y \right] \right\} \bm{)} k_x^2 - y (p_0+i\delta)^2\bm{\big]} ^{\frac{5}{2}}} \nn
		&= \frac{u_0}{4\pi^2 \epsilon} h_v(v) \sigma_x,
		\label{yukawafrompowerlawdisorder}
	\eal
where
	\bal
		&h_v(v)=\int_0^1dx dy \frac{1-\left[ 1-4v^2 x(1-x)\right] y^2}{\sqrt{y(1-y)} \bm{(}  1-y \left\{1-v^2 \left[ 1-(1-2x)^2y \right] \right\} \bm{)} ^{\frac{3}{2}}}
	\eal
is formally convergent for finite $v$, but divergent in the limit $v\rightarrow 0$. Comparing to \eqref{yukawavertexzerov}, we conclude that we need to cut off $h_v$ as soon as it becomes of order $\mathcal{O}(\log\Lambda)$ via $\lim_{v\rightarrow 0} h_v(v) = 4\log\Lambda$.

In contrast to the $\delta$-correlated case, we obtain an $\epsilon$-pole independent of the choice of $v$ and therefore need to include a counter term in principle. Since $v$ turns out to flow to a constant, finite value in the presence of disorder (see Fig.~\ref{flowmodel1logscale}(a)), the contribution from \eqref{yukawafrompowerlawdisorder} is suppressed by factors $1/\log\Lambda$ compared to other terms in the $\beta$-functions and therefore it can be neglected in the flow equations.

\section{Renormalization}
\label{appC}

In the following we use the minimal subtraction scheme, where the divergent parts of the self-energies and vertex corrections are used as counter terms to obtain RG flow equations.
Since we used the definition $\Psib \left( G^{-1}_0 - \Sigma \right) \Psi$ and $\frac{1}{2} \vec{\phi} \left( D^{-1}_0 - \Pi \right) \vec{\phi}$ for the self-energies, we need to add the self-energies and also the counter terms for the interaction vertices to cancel the $\epsilon$-poles. This leads to the renormalized action 
	\begin{equation}
	\begin{aligned}
		&S_{\tn{ren}}= \sum_{n,j,\sigma} \int_k \bmat \Psib^{(1)}_{n,j,\sigma}(k) & \Psib^{(2)}_{n,j,\sigma}(k) \emat \\
		&\bmat \sigma_y (Z_1\omega+i\delta) - Z_2 i \boldsymbol{\sigma}_{\mathbf{z}} \tk - Z_4 i \sigma_x \varepsilon_n(\frac{Z_3}{Z_4}v;\k) & 2 i\ \sgn{\omega} \delta\ \sigma_y \\ 0 & \sigma_y (Z_1 \omega-i\delta) - Z_2i \boldsymbol{\sigma}_{\mathbf{z}}\tk - Z_4 i \sigma_x \varepsilon_n(\frac{Z_3}{Z_4}v;\k) \emat \bmat \Psi^{(1)}_{n,j,\sigma}(k) \\ \Psi^{(2)}_{n,j,\sigma}(k) \emat \\
		&+\frac{1}{2} \int_p \bmat \vec{\phi}^c(p) && \vec{\phi}^q(p) \emat \bmat 0 && 2\left[ Z_5 \left(\omega-i\delta \right)^2-Z_6 a^2\tp^2-c^2 \p^2 \right] \\ 2\left[Z_5  \left( \omega+i\delta \right)^2-Z_6a^2\tp^2-c^2\p^2 \right] && 2i\delta \emat \bmat \vec{\phi}^c(-p) \\ \vec{\phi}^q(-p) \emat \\
		&- iZ_7\frac{g \mu^{\frac{\epsilon}{2}}}{\sqrt{N_f}} \sum_{n,j,\sigma,\sigma'}\int_{k,p} \Psib^{(a)}_{n,j,\sigma}(k+p)\ \Phi^{\alpha}_{\sigma \sigma'}(p) \gamma^\alpha_{(ab)} \sigma_x\ \Psi^{(b)}_{\bar{n},j,\sigma'}(k) \\
	&-Z_8\frac{u_1 \mu^\epsilon}{4} \int_{p_1,p_2,p_3} \tn{Tr} \left\{ \Phi^c (p_1+p_3) \Phi^c(p_2-p_3) + \Phi^q (p_1+p_3) \Phi^q(p_2-p_3) \right\} \tn{Tr} \left\{ \Phi^c (-p_1) \Phi^q(-p_2) \right\} \\
	&-Z_9\frac{u_2 \mu^\epsilon}{4} \int_{p_1,p_2,p_3}  \tn{Tr} \left\{ \left[ \Phi^c (p_1+p_3) \Phi^c(p_2-p_3) + \Phi^q (p_1+p_3) \Phi^q(p_2-p_3)\right] \Phi^c (-p_1) \Phi^q(-p_2) \right\} \\
	&+i Z_{10} \sum_{n,n'}\sum_{j,j'} \sum_{\sigma,\sigma'} \int_{\overset{\vec{k}_1,\vec{k}_2,\vec{k}_3}{\omega_1,\omega_2}} \left[ \Psib^{(a)}_{n,j,\sigma}(\vec{k}_1+\vec{k}_3,\omega_1) \sigma_y \Psib^{(a)}_{n,j,\sigma}(\vec{k}_1,\omega_1) \right] u(\vec{k}_3) \left[ \Psib^{(b)}_{n',j',\sigma'}(\vec{k}_2-\vec{k}_3,\omega_2) \sigma_y \Psib^{(b)}_{n',j',\sigma'}(\vec{k}_2,\omega_2) \right]
	\end{aligned}
	\end{equation}
with the renormalization constants
	\bal
		Z_2 &= 1 - \frac{g^2}{\epsilon}\frac{1}{4\pi^2c} \frac{N_c^2-1}{N_f N_c} h_q(c,v,a),\\
		Z_3 &= 1 + \frac{g^2}{\epsilon}\frac{1}{4\pi^2c} \frac{N_c^2-1}{N_f N_c} h_2(c,v,a),\\
		Z_4 &= 1 - \frac{g^2}{\epsilon}\frac{1}{4\pi^2c} \frac{N_c^2-1}{N_f N_c} h_2(c,v,a),\\
		Z_5 &= 1-\frac{g^2}{\epsilon}\frac{1}{4\pi v }, \\
		Z_6 &= 1-\frac{g^2}{\epsilon}\frac{1}{4\pi a^2 v },\\
		Z_8&= 1+ \frac{u_1}{\epsilon}\frac{N_c^2+7}{2\pi^2 c^2 a} + \frac{u_2}{\epsilon} \frac{2N_c^2-3}{N_c \pi^2 c^2 a}+\frac{u_2^2}{u_1 \epsilon} \frac{3(N_c^2+3)}{2N_c^2 \pi^2 c^2 a},\\
		Z_9&=1+ \frac{u_1}{\epsilon} \frac{6}{\pi^2 c^2 a} + \frac{u_2}{\epsilon}\frac{N_c^2-9}{N_c \pi^2 c^2 a}
	\eal
independent of the choice of the disorder correlation model (i.e.~$\delta$-correlated or power-law), as well as
	\bal
		Z_1 &= 1 - \frac{g^2}{\epsilon} \frac{1}{4\pi^2c} \frac{N_c^2-1}{N_f N_c} h_1(c,v,a) - \frac{u_\delta}{\epsilon} \frac{1}{\pi^2},\\
		Z_7&=1- \frac{g^2}{\epsilon}\frac{1}{8\pi^3 c N_f N_c} h_3(c,v,a),\\
		Z_{10} &= 1-\frac{g^2}{\epsilon} \frac{1}{2\pi^2 c} \frac{N_c^2-1}{N_f N_c} h_1(c,v,a)-\frac{u_\delta}{\epsilon} \frac{2}{\pi^2}
	\eal
for the $\delta$-correlated case and
	\bal
		Z_1 &= 1 - \frac{g^2}{\epsilon} \frac{1}{4\pi^2c} \frac{N_c^2-1}{N_f N_c} h_1(c,v,a) - \frac{u_p}{\epsilon} \frac{1}{\sqrt{1+v^2}\pi^2},\\
		Z_7&=1- \frac{g^2}{\epsilon}\frac{1}{8\pi^3 c N_f N_c} h_3(c,v,a) + \frac{u_p}{\epsilon} \frac{1}{4\pi^2\log\Lambda} h_v(v),\\
		Z_{10} &= 1-\frac{g^2}{\epsilon} \frac{1}{2\pi^2 c} \frac{N_c^2-1}{N_c N_f} h_1(c,v,a)-\frac{u_p}{\epsilon} \frac{2}{\sqrt{1+v^2}\pi^2}
	\eal
for the power-law correlated case. For the following calculations, we will use the notation $Z_i = 1+ \frac{g^2}{\epsilon} Z_{i1,g}+ \frac{u_{\delta/p}}{\epsilon} Z_{i1,u}$ except for $Z_8$ and $Z_9$, where we write $Z_{i} = 1+ \frac{u_1}{\epsilon}Z_{i1,u_1}+\frac{u_2}{\epsilon}Z_{i1,u_2}+\frac{u_2^2}{u_1 \epsilon}Z_{i1,u_1u_2}$.

Defining the bare momenta and fields as
	\begin{equation}
	\begin{aligned}
		&\omega_B= Z_1 Z_4^{-1}\omega,\quad \tk_B = Z_2 Z_4^{-1} \tk,\quad \k_B=\k, \\
		&\Psi_B = Z_1^{-1/2} Z_2^{(2-d)/2} Z_4^{d/2} \Psi,\quad \vec{\phi}_B = Z_1^{-3/2}  Z_2^{(2-d)/2} Z_4^{(d+1)/2} Z_5^{1/2} \vec{\phi},
	\end{aligned}
	\end{equation}
the action can be brought into its original form by defining the bare velocities and coupling constants via
	\begin{equation}
	\begin{aligned}
		&c_B = Z_1 Z_4^{-1} Z_5^{-1/2} c,\quad v_B = Z_3 Z_4^{-1} v, \quad a_B = Z_1 Z_2^{-1} Z_5^{-1/2} Z_6^{1/2} a,\\
		&g_B = \mu^{\epsilon/2} Z_1^{1/2} Z_2^{(2-d)/2} Z_4^{(d-5)/2} Z_5^{-1/2} Z_7\ g, \quad u_{\delta/p,B} = \mu^\epsilon Z_2^{2-d} Z_4^{d-4} Z_{10}\ u_{\delta/p}, \\
		&u_{1,B} = \mu^\epsilon Z_1^3 Z_2^{2-d} Z_4^{d-5} Z_5^{-2} Z_8\ u_1, \quad u_{2,B} = \mu^\epsilon Z_1^3 Z_2^{2-d} Z_4^{d-5} Z_5^{-2} Z_9\ u_2.
	\end{aligned}
	\end{equation}
During the renormalization procedure we absorbed some renormalization factors $Z_i$ into the infinitesimal $\delta=0^+$ since the $Z_i$ are formally of order $\mathcal{O}(1)$ in the correct order of limits, namely expanding in the interactions first and in small $\epsilon=3-d$ second.

The $\beta$-functions of the parameters $i \in \left\{ c,v,a,g,u_1,u_2,u_{\delta/p} \right\}$ are obtained from the coupled equations
	\bal
		\beta_c&=\mu \frac{dc}{d\mu} = c\sum_i \beta_i \left( - \frac{\partial_i Z_1}{Z_1}+\frac{\partial_i Z_4}{Z_4}+\frac{1}{2} \frac{\partial_i Z_5}{Z_5} \right),\\
		\beta_v &=v \sum_i \beta_i \left( \frac{\partial_i Z_4}{Z_4} - \frac{\partial_i Z_3}{Z_3} \right), \\
		\beta_a &= a \sum_i \beta_i \left( -\frac{\partial_i Z_1}{Z_1}+\frac{\partial_i Z_2}{Z_2} + \frac{1}{2} \frac{\partial_i Z_5}{Z_5} - \frac{1}{2} \frac{\partial_i Z_6}{Z_6} \right), \\
		\beta_g &= -\frac{\epsilon}{2} g+g \sum_i \beta_i \bigg( - \frac{1}{2} \frac{\partial_i Z_1}{Z_1}+ \frac{d-2}{2} \frac{\partial_i Z_2}{Z_2} + \frac{5-d}{2} \frac{\partial_i Z_4}{Z_4}+ \frac{1}{2} \frac{\partial_i Z_5}{Z_5} - \frac{\partial_i Z_7}{Z_7} \bigg), \\
		 \beta_{u_1} &= -\epsilon u_1 + u_1 \sum_i \beta_i \bigg[ -3 \frac{\partial_i Z_1}{Z_1}+(d-2) \frac{\partial_i Z_2}{Z_2} + (5-d) \frac{\partial_i Z_4}{Z_4} + 2 \frac{\partial_i Z_5}{Z_5} - \frac{\partial_i Z_8}{Z_8} \bigg], \\
		 \beta_{u_2}&=-\epsilon u_2 + u_2 \sum_i \beta_i \bigg[ -3 \frac{\partial_i Z_1}{Z_1}+(d-2) \frac{\partial_i Z_2}{Z_2} + (5-d) \frac{\partial_i Z_4}{Z_4} + 2 \frac{\partial_i Z_5}{Z_5} - \frac{\partial_i Z_9}{Z_9} \bigg], \\
		\beta_{u_{\delta/p}} &=-\epsilon u_{\delta/p} + u_{\delta/p} \sum_i \beta_i \bigg[ (d-2) \frac{\partial_i Z_2}{Z_2} + (4-d) \frac{\partial_i Z_4}{Z_4} - \frac{\partial_i Z_{10}}{Z_{10}} \bigg]
	\eal
and are solved to give
	\bal
		\beta_c &= \frac{1}{2} g^2 c \left( 2 Z_{11,g} - 2 Z_{41,g} - Z_{51,g} \right) + c u_{\delta/p} Z_{11,u}, \\
		\beta_v &= g^2 v \left( Z_{31,g} - Z_{41,g} \right), \\
		\beta_a &= \frac{1}{2} g^2 a \left( 2 Z_{11,g}-2 Z_{21,g} - Z_{51,g} + Z_{61,g} \right) + a u_{\delta/p} Z_{11,u}, \\
		\beta_g &= -\frac{\epsilon}{2} g + \frac{1}{2} g^3 \left( Z_{11,g} - Z_{21,g} - 2 Z_{41,g} -Z_{51,g}+ 2 Z_{71,g} \right)+\frac{1}{2} g u_{\delta/p} \left( Z_{11,u} + 2 Z_{71,u} \right), \\
		\beta_{u_1} &= -\epsilon u_1 +u_1^2 Z_{81,u_1} + u_1 u_2 Z_{81,u_2} + u_2^2 Z_{81,u_1u_2} + u_1 g^2 \left( 3 Z_{11,g} - Z_{21,g} - 2 Z_{41,g} - 2 Z_{51,g} \right) \nn
		&+ 3 u_1 u_{\delta/p} Z_{11,u}, \\
		\beta_{u_2} &= - \epsilon u_2 + u_2^2 Z_{91,u_2} + u_1 u_2 Z_{91,u_1} + u_2 g^2 \left( 3Z_{11,g} - Z_{21,g} - 2 Z_{41,g} - 2 Z_{51,g} \right) + 3 u_2 u_{\delta/p} Z_{11,u}, \\
		\beta_{u_{\delta/p}} &= -\epsilon u_{\delta/p} + u^2_{\delta/p} Z_{101,u}+ u_{\delta/p} g^2 \left( - Z_{21,g} -  Z_{41,g} + Z_{101,g} \right)
		\label{appendixbetafunctions}
	\eal
to leading order in $\epsilon$.

The general scaling form of the correlation functions
	\bal
		&G^{(m,m,2n)}(\{ k_i \}, \mu, c,v,a,g,u_1,u_2,u_{\delta/p}) = \delta^{(d+1)}( \{ k_i \} ) \langle \Psi(k_1) ... \Psi(k_m) \Psib(k_{m+1})... \Psib(k_{2m}) \phi(k_{2m+1}) ... \phi(k_{2m+2n}) \rangle,
	\eal
with the $\delta$-function ensuring energy and momentum conservation, can be derived from the RG equation
\endgroup
	\bal
		&\left\{ \sum_{j=1}^{2m+2n} \left( z_\omega \omega_j \partial_{\omega_j} + z_{\tk} \tk_j \nabla_{\tk_j} + \k_j \nabla_{\k_j} \right) - \sum_i \beta_i \partial_i - 2m \left( \eta_\Psi-\frac{d+2}{2} \right) - 2n \left( \eta_\phi - \frac{d+3}{2} \right) - \left[ z_\omega+ z_\tk (d-2) + 2 \right] \right\} \nn
		& G^{(m,m,2n)}(\{ k_i \}, \mu, c,v,a,g,u_1,u_2,u_{\delta/p}) =0.
	\eal
\begingroup
\allowdisplaybreaks
Here we defined the dynamical critical exponents 
	\bal
		z_\omega &= 1 + \frac{d \ln (Z_1/Z_4)}{d\ln \mu} = 1-g^2 \left( Z_{11,g} - Z_{41,g} \right) - u_{\delta/p} Z_{11,u}, \\
		z_\tk &= 1 + \frac{d \ln (Z_2/Z_4)}{d\ln \mu} = 1-g^2 \left( Z_{21,g} - Z_{41,g} \right)
		\label{criticalexponents}
	\eal
and the anomalous dimensions of the fields
	\bal
		\eta_\Psi &= \frac{1}{2} \frac{d\ln Z_\Psi}{d\ln \mu} = \frac{g^2}{2} \left( Z_{11,g} + Z_{21,g} -3 Z_{41,g} \right) +\frac{u_{\delta/p}}{2} Z_{11,u} \\
		\eta_\phi &= \frac{1}{2} \frac{d\ln Z_\phi}{d\ln \mu} = \frac{g^2}{2} \left( 3 Z_{11,g} + Z_{21,g} -4 Z_{41,g} - Z_{51,g} \right) + \frac{3}{2} u_{\delta/p} Z_{11,u}
		\label{anomalousdimensions}
	\eal
where $Z_\Psi = Z_1^{-1} Z_2^{2-d} Z_4^d$ and $Z_\phi = Z_1^{-3} Z_2^{2-d} Z_4^{d+1} Z_5$.

At fixed points, where the $\beta$-functions vanish, the fermionic and bosonic propagators take the scaling form
	\bal
		G(k) &= \frac{1}{\av{k_y}^{1-2\tilde{\eta}_\Psi}} f_{\Psi} \left(\frac{\omega}{\av{k_y}^{z_\omega}}, \frac{\tk}{\av{k_y}^{z_\tk}}, \frac{k_x}{\av{k_y}} \right), \\
		D(k) &= \frac{1}{\av{k_y}^{2-2\tilde{\eta}_\phi}} f_{\phi} \left(\frac{\omega}{\av{k_y}^{z_\omega}}, \frac{\tk}{\av{k_y}^{z_\tk}}, \frac{k_x}{\av{k_y}} \right),
	\eal
where $f_{\Psi/\phi}$ are universal scaling functions and
	\bal
		\tilde{\eta}_{\Psi/\phi}&= \eta_{\Psi/\phi} + \frac{z_\omega+z_\tk(1-\epsilon) -2+\epsilon}{2} 
	\eal
are effective anomalous dimensions.

The flow of the parameters is usually parametrized in terms of a logarithmic length scale $\ell = \log(\mu_0/\mu)$, where e.g. $\frac{dg}{d\ell} = - \beta_g$ describes the flow to lower energies for increasing $\ell$. In the context of this paper, it is convenient to introduce a different flow parameter to be able to follow the flow up to arbitrary strong disorder. Since $\beta_{u_{\delta/p}}$ is positive definite (at least when the flow to strong disorder begins), we choose $\log u_{\delta/p}$ as an alternative flow parameter. Furthermore, we introduce new coupling constants $\tilde{g} = g^2/u_{\delta/p}$, $\tilde{u}_1 = u_1/u_{\delta/p}$ and $\tilde{u}_2 = u_2/u_{\delta/p}$, whose flow equations can be written down directly, e.g.
	\bal
		\frac{d\tilde{g}}{d\log u_{\delta/p}} = \frac{dg^2}{du_{\delta/p}} - \frac{g^2}{u_{\delta/p}} = \frac{2g \beta_g}{\beta_{u_{\delta/p}}} - \tilde{g}.
	\eal
Using the explicit expressions of the $\beta$-functions in \eqref{appendixbetafunctions}, we can simplify
	\bal
		\frac{2g \beta_g}{\beta_{u_{\delta/p}}} = \tilde{g} \frac{-\frac{\epsilon}{u_{\delta/p}} + \tilde{g} \left( Z_{11,g} - Z_{21,g} - 2 Z_{41,g}-Z_{51,g}+ 2 Z_{71,g} \right) + Z_{11,u} + 2 Z_{71,u} }{-\frac{\epsilon}{u_{\delta/p}} + Z_{101,u}+ \tilde{g} \left( - Z_{21,g} - Z_{41,g} + Z_{101,g} \right)},
	\eal
where we can drop the $\frac{\epsilon}{u_{\delta/p}}$-terms for $u_{\delta/p} \rightarrow \infty$. The flow equations of all parameters in terms of the RG scale $\log u_{\delta/p}$ can then be derived to be
	\bal
		\frac{d\tilde{g}}{d\log u_{\delta/p}} &= \tilde{g} \left( \frac{\tilde{g} \left( Z_{11,g} - Z_{21,g} - 2 Z_{41,g}-Z_{51,g}+ 2 Z_{71,g} \right)+ Z_{11,u} + 2 Z_{71,u}}{ Z_{101,u}+ \tilde{g} \left( - Z_{21,g} - Z_{41,g} + Z_{101,g} \right)} -1 \right), \\
		\frac{d\tilde{u}_1}{d\log u_{\delta/p}} &= \tilde{u}_1 \left( \frac{\tilde{u}_1 Z_{81,u_1} + \tilde{u}_2 Z_{81,u_2} + \frac{\tilde{u}_2^2}{\tilde{u}_1} Z_{81,u_1u_2}+ \tilde{g} \left( 3 Z_{11,g} - Z_{21,g} - 2 Z_{41,g} - 2 Z_{51,g} \right) + 3Z_{11,u}}{Z_{101,u}+ \tilde{g} \left( - Z_{21,g} -  Z_{41,g} + Z_{101,g} \right)} -1\right), \\
		\frac{d\tilde{u}_2}{d\log u_{\delta/p}} &= \tilde{u}_2 \left( \frac{ \tilde{u}_2 Z_{91,u_2} + \tilde{u}_1 Z_{91,u_1} + \tilde{g} \left( 3Z_{11,g} - Z_{21,g} - 2 Z_{41,g} - 2 Z_{51,g} \right) + 3 Z_{11,u}}{Z_{101,u}+ \tilde{g} \left( - Z_{21,g} - Z_{41,g} + Z_{101,g} \right)} -1 \right),\\
		\frac{dc}{d\log u_{\delta/p}}&= c \frac{\frac{1}{2} \tilde{g} \left( 2 Z_{11,g} - 2 Z_{41,g} - Z_{51,g} \right) + Z_{11,u}}{Z_{101,u}+ \tilde{g} \left( - Z_{21,g} - Z_{41,g} + Z_{101,g} \right)}, \\
		\frac{dv}{d\log u_{\delta/p}}&= v \frac{\tilde{g} \left( Z_{31,g} - Z_{41,g} \right)}{Z_{101,u}+ \tilde{g} \left( - Z_{21,g} -  Z_{41,g} + Z_{101,g} \right)},\\
		\frac{da}{d\log u_{\delta/p}}&= a \frac{\frac{1}{2} \tilde{g} \left( 2 Z_{11,g}-2 Z_{21,g} - Z_{51,g} + Z_{61,g} \right) + Z_{11,u}}{Z_{101,u}+ \tilde{g} \left( - Z_{21,g} - Z_{41,g} + Z_{101,g} \right)}.
	\eal
\endgroup

\twocolumngrid

\end{document}